\documentclass[aps,prb,twocolumn,groupedaddress]{revtex4-1}

\usepackage[utf8]{inputenc}
\usepackage[]{amsmath}
\usepackage{amsfonts}
\usepackage[]{color}
\usepackage[english]{babel}

\usepackage{bbm}
\usepackage{subcaption}
\usepackage{tikz}
\usepackage{tikz-feynman}
\usepackage{pgfplots}

\pgfplotsset{width=6.8cm}
\pgfplotsset{compat=1.3}
\renewcommand{\vec}[1]{\boldsymbol #1}
\newcommand{\h}{\hspace{1pt}}
\newcommand{\mh}{\hspace{-1pt}}
\newcommand{\hh}{\hspace{0.5pt}}
\newcommand{\mhh}{\hspace{-0.5pt}}

\usepackage{hyperref}

\graphicspath{{./plots_sub/}}

\begin{document}

\tikzset{->-/.style={decoration={
  markings,
  mark=at position #1 with {\arrow{Latex[scale=1.6]}}},postaction={decorate}}}


\title{Truncated-Unity Parquet Equations: \\ Application to the Repulsive Hubbard Model}


\author{C. J. Eckhardt$^{1}$}
\email[]{christian.eckhardt@rwth-aachen.de}
\author{G. A. H. Schober$^{1}$}
\author{J. Ehrlich$^{1,2}$}
\author{C. Honerkamp$^{1,3}$}
\affiliation{$^{1}$ Institute for Theoretical Solid State Physics, RWTH Aachen University, 52074 Aachen,
Germany 
\\ $^{2}$ Peter Gr\"unberg Institute and Institute for Advanced Simulation, Forschungszentrum 52435 J\"ulich, Germany \\
$^{3}$ JARA-FIT, J\"ulich Aachen Research Alliance - Fundamentals of Future Information Technology, Germany }


\date{\today}

\begin{abstract}
The parquet equations are a self-consistent set of equations for the effective two-particle vertex of an interacting many-fermion system. 
The application of these equations to bulk models is, however, demanding due to the complex emergent momentum and frequency structure of the vertex.
Here, we show how a channel-decomposition by means of truncated unities, which was developed in the context of the functional renormalization group to efficiently treat the momentum dependence, can be transferred to the parquet equations.
This leads to a significantly reduced complexity and memory consumption scaling only linearly with the number of discrete momenta.
We apply this technique to the half-filled repulsive Hubbard model on the square lattice and present approximate solutions for the channel-projected vertices and the full reducible vertex.
\end{abstract}

\pacs{}

\maketitle

\section{Introduction}
The parquet approach was introduced more than five decades ago as a method to analyze interacting many-fermion systems \cite{sudakov,dedominicis,bickers}.
Since then, it has helped significantly in understanding the physics of magnetic impurities in metals \cite{abrikosov} as well as the breakdown of Fermi liquid behavior in one-dimensional metals \cite{dzyaloshinskii,solyom}.
For a recent application to Hubbard nanorings, see Ref.~\onlinecite{valli}.
A main advantage of the parquet scheme is that it can be made self-consistent at the single-particle and two-particle level \cite{Yang09,rohringer17}.
It has been shown that the parquet approximation (to be introduced later) is a thermodynamically consistent and conserving, \(\Phi\)-derivable approximation in Baym's sense at the one-particle level \cite{Fabian2018}.
From early on, it has been clear that the parquet approximation is closely related to perturbative renormalization group (RG) schemes.
At least regarding the modern fermionic functional renormalization group (fRG) flavors (for reviews, see e.g.~Refs.~\onlinecite{metzner2012,platt13}), it has been understood that the parquet approximation and the fRG in the usual truncations sum the same classes of diagrams \cite{kugler}.
However, it has also been known that in the fRG, due to the unavoidable truncation of the hierarchy of flow equations, certain combinations of internal lines are suppressed compared to the contributions kept in the parquet approximation.
Parts of these missing contributions can be recollected by refined flow equations \cite{katanin2009,maier2012,eberlein2014}.
Quite recently, a systematic multi-loop fRG scheme was proposed \cite{kugler} that in simplified models reconciles the fRG results with those of the parquet approximation in a quantitative manner.
These attempts to lift the fRG on higher levels already indicate that fRG approaches may offer advantages despite the fact that they do not readily contain the full perturbative corrections of the parquet approximation.
Indeed, considering the vast more recent literature on standard zero- to two-dimensional correlated many-fermion lattice systems, the applications of RG schemes seem to outnumber clearly those of parquet schemes.

In order to make the comparison more specific, we mention two recent state-of-the-art parquet studies of the two-dimensional Hubbard model \cite{jarrell_sym,held_new}. 
In these works, the finest momentum resolution reaches 6$\times $6 due to memory constraints.
Hence, very few points are located in the vicinity of the Fermi surfaces, and many interesting questions like the generation of unconventional superconductivity, the opening of a pseudogap, or the tendency toward incommensurate or stripe ordering are hard to study. By contrast, in the fRG approaches, $\mathcal O(100)$ momentum-space patches were employed at an early stage without the use of parallel computers \cite{honerkampepjb2001}, and 14$\times $14 grids were also analyzed \cite{honerkamp2004}.
It should be mentioned that these works did not consider the frequency dependence of the interactions, which is usually kept in the parquet studies. Yet, more recent fRG schemes are about to remedy this shortcoming while still reaching similar momentum resolutions.

A substantial progress in simplifying the description of the momentum structure has been provided by channel-decomposed fRG schemes \cite{husemann,wang,julian}.
The main simplification there consists in expressing the effective two-particle interaction, which depends on three momenta (usually two incoming and one outgoing momentum, the fourth one being fixed by momentum conservation), by three interaction functions that describe the interaction between specific fermion bilinears in particle-particle and particle-hole channels. Each of these interaction functions depends strongly on one `bosonic' momentum but only weakly on two other wavevectors, hence the latter dependencies can be expanded in a suitable form-factor basis. In the simplest cases, this means that the interacting fermion bilinears live on nearby sites on the lattice, while longer bilinears are usually not relevant.
This then allows for a well-convergent and physically meaningful truncation of the form-factor expansion. Without frequency dependence, the channel-decomposed fRG schemes can be parallelized efficiently and pushed to very fine momentum resolutions with thousands of momenta in the Brillouin zone, in conjunction with convergence checks in the form-factor truncation \cite{julian,David17}.
Currently, the inclusion of the frequency dependence and self-energy effects into these schemes is on the way (see e.g.~Refs.~\onlinecite{giering,eberlein15,vilardi}).
For the frequency dependence, a channel decomposition was also shown to yield meaningful results for impurity models \cite{karrasch}, but in general, the complex frequency structure requires a more sophisticated description \cite{wentzell,vilardi}.

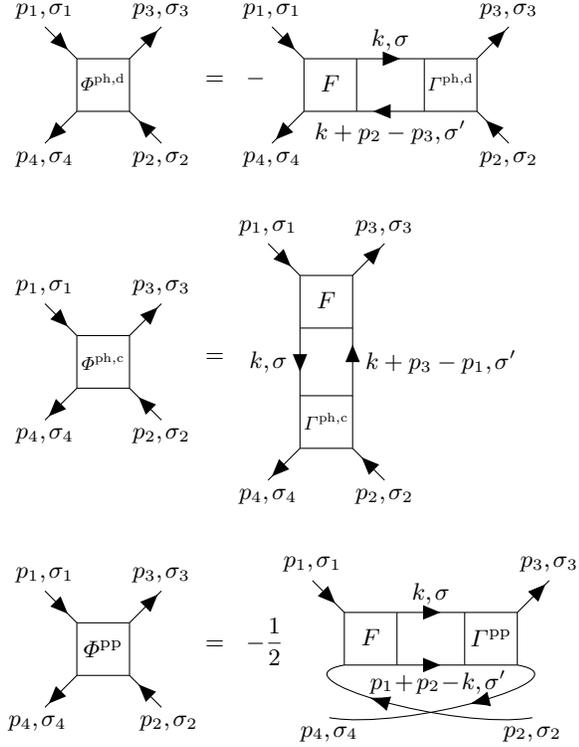
\begin{figure}[t]
\begin{align*}
\begin{tikzpicture}[baseline=(current bounding box.center)]
	\begin{feynman}
		\vertex (a);
		\vertex [right= 0.7cm of a] (b);
		\vertex [below= 0.7cm of a] (c);
		\vertex [right= 0.7cm of c] (d);
		\vertex [dot,label=\(\mbox{\scalebox{0.8}{$\varPhi^{\mathrm{ph,d}}$}}\),below right=0.54cm and 0.35cm of a] (e);
		\vertex [label= \(p_1{,} \h \sigma_1\) ,above left=0.6cm of a] (f);
		\vertex [label= \(p_3{,} \h \sigma_3\),above right=0.6cm of b] (g);
		\vertex [label=270: \(p_4{,} \h \sigma_4\),below left=0.6cm of c] (h);
		\vertex [label=270:\(p_2{,} \h \sigma_2\),below right=0.6cm of d] (i);
		\diagram*[baseline = (e)]{
			(a) -- (b)
			(a) -- (c)
			(b) -- (d)
			(c) -- (d)
			(f) --[fermion] (a)
			(g) --[anti fermion] (b)			
			(h) --[anti fermion] (c)
			(i) --[fermion] (d)
		};
	\end{feynman}
\end{tikzpicture}
& = \hspace{2mm} - \hspace{-4mm}
\begin{tikzpicture}[baseline=(current bounding box.center)]
	\begin{feynman}
		\vertex (a);
		\vertex [right= 0.7cm of a] (b);
		\vertex [below= 0.7cm of a] (c);
		\vertex [right= 0.7cm of c] (d);
		\vertex [right= 0.9cm of b] (j);
		\vertex [right= 0.7cm of j] (k);
		\vertex [below= 0.7cm of j] (l);
		\vertex [right= 0.7cm of l] (m);
		\vertex [label= \(p_1{,} \h \sigma_1\),above left=0.6cm of a] (f);
		\vertex [label= \(p_3{,} \h \sigma_3\),above right=0.6cm of k] (g);
		\vertex [label=270: \(p_4{,} \h \sigma_4\),below left=0.6cm of c] (h);
		\vertex [label=270:\(p_2{,} \h \sigma_2\),below right=0.6cm of m] (i);
		\vertex [dot,label=\(F\),below right=0.55cm and 0.35cm of a] (e);
		\vertex [dot,label=\(\mbox{\scalebox{0.8}{$\varGamma^{\mathrm{ph,d}}$}}\),below right=0.54cm and 0.35cm of j] (n);
		\diagram*[baseline = (e)]{
			(a) -- (b)
			(a) -- (c)
			(b) -- (d)
			(c) -- (d)
			(j) -- (k)
			(j) -- (l)
			(l) -- (m)
			(k) -- (m)			
			(f) --[fermion] (a)
			(g) --[anti fermion] (k)			
			(h) --[anti fermion] (c)
			(i) --[fermion] (m)
			(b) --[fermion,edge label=\(k{,} \h \sigma\)] (j)
			(d) --[anti fermion,edge label'=\(k+p_2 - p_3{,} \h \sigma'\)] (l)
		};
	\end{feynman}
\end{tikzpicture} \\[10pt]
\begin{tikzpicture}[baseline=(current bounding box.center)]
	\begin{feynman}
		\vertex (a);
		\vertex [right= 0.7cm of a] (b);
		\vertex [below= 0.7cm of a] (c);
		\vertex [right= 0.7cm of c] (d);
		\vertex [dot,label=\(\mbox{\scalebox{0.8}{$\varPhi^{\mathrm{ph,c}}$}}\),below right=0.57cm and 0.35cm of a] (e);
		\vertex [label= \(p_1{,} \h \sigma_1\),above left=0.6cm of a] (f);
		\vertex [label= \(p_3{,} \h \sigma_3\),above right=0.6cm of b] (g);
		\vertex [label=270: \(p_4{,} \h \sigma_4\),below left=0.6cm of c] (h);
		\vertex [label=270:\(p_2{,} \h \sigma_2\),below right=0.6cm of d] (i);
		\diagram*[baseline = (e)]{
			(a) -- (b)
			(a) -- (c)
			(b) -- (d)
			(c) -- (d)
			(f) --[fermion] (a)
			(g) --[anti fermion] (b)			
			(h) --[anti fermion] (c)
			(i) --[fermion] (d)
		};
	\end{feynman}
\end{tikzpicture}
& =
\begin{tikzpicture}[baseline=(current bounding box.center)]
	\begin{feynman}
		\vertex (a);
		\vertex [right= 0.7cm of a] (b);
		\vertex [below= 0.7cm of a] (c);
		\vertex [right= 0.7cm of c] (d);
		\vertex [below= 0.9cm of c] (j);
		\vertex [right= 0.7cm of j] (k);
		\vertex [below= 0.7cm of j] (l);
		\vertex [right= 0.7cm of l] (m);
		\vertex [label= \(p_1{,} \h \sigma_1\),above left=0.6cm of a] (f);
		\vertex [label= \(p_3{,} \h \sigma_3\),above right=0.6cm of b] (g);
		\vertex [label=270: \(p_4{,} \h \sigma_4\),below left=0.6cm of l] (h);
		\vertex [label=270:\(p_2{,} \h \sigma_2\),below right=0.6cm of m] (i);
		\vertex [dot,label=\(F\),below right=0.55cm and 0.35cm of a] (e);
		\vertex [dot,label=\(\mbox{\scalebox{0.8}{$\varGamma^{\mathrm{ph,c}}$}}\),below right=0.57cm and 0.35cm of j] (n);	
		\diagram*[baseline = (e)]{
			(a) -- (b)
			(a) -- (c)
			(b) -- (d)
			(c) -- (d)
			(j) -- (k)
			(j) -- (l)
			(l) -- (m)
			(k) -- (m)			
			(f) --[fermion] (a)
			(g) --[anti fermion] (b)			
			(h) --[anti fermion] (l)
			(i) --[fermion] (m)
			(c) --[fermion,edge label'=\(k{,} \h \sigma\,\)] (j)
			(d) --[anti fermion,edge label=\(\,k+p_3-p_1{,} \h \sigma'\)] (k)		
		};
	\end{feynman}
\end{tikzpicture} \\[10pt]
\begin{tikzpicture}[baseline=(current bounding box.center)]
	\begin{feynman}
		\vertex (a);
		\vertex [right= 0.7cm of a] (b);
		\vertex [below= 0.7cm of a] (c);
		\vertex [right= 0.7cm of c] (d);
		\vertex [dot,label=\(\varPhi^{\mathrm{pp}}\),below right=0.57cm and 0.35cm of a] (e);
		\vertex [label= \(p_1{,} \h \sigma_1\),above left=0.6cm of a] (f);
		\vertex [label= \(p_3{,} \h \sigma_3\),above right=0.6cm of b] (g);
		\vertex [label=270: \(\hspace{-2mm}p_4{,} \h \sigma_4\),below left=0.6cm of c] (h);
		\vertex [label=270:\(p_2{,} \h \sigma_2 \hspace{-2mm}\),below right=0.6cm of d] (i);
		\diagram*[baseline = (e)]{
			(a) -- (b)
			(a) -- (c)
			(b) -- (d)
			(c) -- (d)
			(f) --[fermion] (a)
			(g) --[anti fermion] (b)			
			(h) --[anti fermion] (c)
			(i) --[fermion] (d)
		};
	\end{feynman}
\end{tikzpicture}
& = 
\hspace{2mm}- \frac{1}{2}\hspace{-2mm}
\begin{tikzpicture}[baseline=(current bounding box.center)]
	\begin{feynman}
		\vertex (a);
		\vertex [right= 0.7cm of a] (b);
		\vertex [below= 0.7cm of a] (c);
		\vertex [right= 0.7cm of c] (d);
		\vertex [right= 0.9cm of b] (j);
		\vertex [right= 0.7cm of j] (k);
		\vertex [below= 0.7cm of j] (l);
		\vertex [right= 0.7cm of l] (m);	
		\vertex [label= \(p_1{,} \h \sigma_1\),above left=0.6cm of a] (f);
		\vertex [label= \(p_3{,} \h \sigma_3\),above right=0.6cm of k] (g);
		\vertex [label=270: \(p_4{,} \h \sigma_4\),below left=0.7cm and 0.2cm of c] (h);
		\vertex [label=270:\(p_2{,} \h \sigma_2\),below right=0.7cm and 0.2cm of m] (i);
		\vertex [label=270:\(p_1\!+\mh p_2\!-\!k{,} \h \sigma'\),above right= 0.05cm and 0.55cm of d];
		\vertex [dot,label=\(F\),below right=0.55cm and 0.35cm of a] (e);
		\vertex [dot,label=\(\varGamma^{\mathrm{pp}}\),below right=0.57cm and 0.35cm of j] (n);		
		\draw[->-=0.4] (m) to [out=-10,in=-10] (h);		
		\draw[->-=0.7] (i) to [out=190,in=-170] (c);		
		\diagram*[baseline = (e)]{
			(a) -- (b)
			(a) -- (c)
			(b) -- (d)
			(c) -- (d)
			(j) -- (k)
			(j) -- (l)
			(l) -- (m)
			(k) -- (m)										
			(f) --[fermion] (a)
			(g) --[anti fermion] (k)		
			(b) --[fermion,edge label=\(k{,} \h \sigma\)] (j)
			(d) --[fermion] (l)			
		};
	\end{feynman}
\end{tikzpicture}
\end{align*}
\caption{Graphical representation of parquet equations. \label{graphical_parquet}}
\end{figure}

The main goal of the present article is to show how the advantages of the channel decomposition explored in fRG studies can be transferred to the parquet equations.
We show that without the frequency dependence, this readily gives meaningful results with a high momentum resolution. In particular, the channel decomposition of the two-particle interaction reduces the memory required for numerically evaluating the parquet equations from $\mathcal O(N^3)$ to $\mathcal O(N)$, where $N$ is the number of momenta in the first Brillouin zone.
This makes is plausible that the channel decomposition will also be beneficial in cases where frequency-dependent interactions are considered.

The article is organized as follows. We begin in Sct.~\ref{sec:channel_decomposition} by briefly describing the parquet equations. After that, we introduce the projections onto the direct particle-hole, crossed particle-hole and particle-particle channel, and subsequently derive the truncated-unity (TU) parquet equations. In Sct.~\ref{sec:cross}, we further derive the cross projections between the different channels, which are necessary for iteratively solving the TU parquet equations. In Sct.~\ref{sec:numerical}, we discuss the advantages of the channel decomposition with respect to computational complexity and memory cost. Following this, we provide more details of our numerical implementation in Sct.~\ref{sec:numerical_implementation}, and we present our results for the half-filled repulsive Hubbard model in Sct.~\ref{sec:results}. Finally, the appendix is concerned with the derivation of channel-decomposed parquet equations for general spin-SU(2)-symmetric systems.

\section{Channel decomposition} \label{sec:channel_decomposition}

We start from the parquet equations as described, for example, in Ref.~\onlinecite{held}. 
These are formulated in terms of the full Green function~$G$, the full (one-particle-irreducible) two-particle-reducible vertex $F$, 
the (in channel $r$ two-particle-) reducible vertices $\varPhi^{r}$ in the direct particle-hole channel ($r = \textnormal{ph,d}$),
the crossed particle-hole channel ($r = \textnormal{ph,c}$), and the particle-particle channel ($r = \textnormal{pp}$),
as well as the corresponding (in channel $r$ two-particle-) irreducible vertices $\varGamma^{r}$ in each channel.
Explicitly, the parquet equations read as follows:\footnote{We mainly follow the conventions of Refs.~\onlinecite{Schober, Schober17, held}; however, the two-point Green function used here contains an additional factor $\beta = 1/(k_{\mathrm B} T)$, and all four-point functions contain an additional factor $2$ compared to those used in Refs.~\onlinecite{Schober, Schober17}. Further note that the functions $\varPhi^{r}$ on the left-hand side of the parquet equations are not identical to the corresponding functions used in the fRG framework.}
\begin{widetext}
\begin{align}
 \varPhi_{\sigma_1 \sigma_2 \sigma_3 \sigma_4}^{\mathrm{ph,d}}(p_1, p_2, p_3)
 & = -\frac{k_{\mathrm B} T}{N} \sum_{k, \h \sigma, \h \sigma'} F_{\sigma_1 \sigma' \sigma \sigma_4}(p_1, k + p_2 - p_3, k) \, G(k) \, G(k + p_2 - p_3) \, \varGamma^{\mathrm{ph,d}}_{\sigma \sigma_2 \sigma_3 \sigma'}(k, p_2, p_3) \,, \label{eq:parquet1_full} \\[5pt]
 \varPhi_{\sigma_1 \sigma_2 \sigma_3 \sigma_4}^{\mathrm{ph,c}}(p_1, p_2, p_3)
 & = \frac{k_{\mathrm B} T}{N} \sum_{k, \h \sigma, \h \sigma'} F_{\sigma_1 \sigma' \sigma_3 \sigma}(p_1, k + p_3 - p_1, p_3) \, G(k) \, G(k + p_3 - p_1) \, \varGamma^{\mathrm{ph,c}}_{\sigma \sigma_2 \sigma' \sigma_4}(k, p_2, k + p_3 - p_1) \,, \label{eq:parquet2_full} \\[5pt]
 \varPhi_{\sigma_1 \sigma_2 \sigma_3 \sigma_4}^{\mathrm{pp}}(p_1, p_2, p_3)
 & = -\frac 1 2 \h \frac{k_{\mathrm B} T}{N} \sum_{k, \h \sigma, \h \sigma'} F_{\sigma_1 \sigma_2 \sigma \sigma'}(p_1, p_2, k) \, G(k) \, G(p_1 + p_2 - k) \, \varGamma^{\mathrm{pp}}_{\sigma \sigma' \sigma_3 \sigma_4}(k, p_1 + p_2 - k, p_3) \,. \label{eq:parquet3_full}
\end{align}
\end{widetext}
Here, the function arguments $k \equiv (\omega, \h \vec k) \equiv (k_0, \h \vec{k})$ are multi-indices comprising fermionic Matsubara frequencies (at temperature $T$) and Bloch momenta (of which there are $N$ in the first Brillouin zone).
Each vertex in Eqs.~\eqref{eq:parquet1_full}--\eqref{eq:parquet3_full} depends on only three momenta/frequencies, since the fourth is always determined by momentum/energy conservation.
Furthermore, each vertex depends on four spin indices, while we have assumed the full Green function to be spin independent corresponding to the SU(2)-symmetric case. The different vertices appearing in Eqs.~\eqref{eq:parquet1_full}--\eqref{eq:parquet3_full} are related through
\begin{equation} \label{eq:Fsum}
F = \varLambda + \varPhi^{\mathrm{ph,d}} + \varPhi^{\mathrm{ph,c}} + \varPhi^{\mathrm{pp}} \,,
\end{equation}
where $\varLambda$ denotes the fully (two-particle-) irreducible vertex,
and by
\begin{equation} \label{eq:Fsplit}
\varGamma^{r} = F - \varPhi^{r} \,.
\end{equation}
Throughout this article, we will use the {\itshape parquet approxi\-{}mation} \cite{held}, which identifies the fully irreducible vertex with the initial interaction given in the Hubbard model by Eq.~\eqref{ini_int}, hence
$\Lambda \equiv F^0$.
Graphically, the parquet equations can be represented by means of Feynman diagrams as in Fig.~\ref{graphical_parquet} (see also Ref.~\onlinecite{held}).

Next, we define the bosonic Matsubara frequencies and corresponding 
transfer ($\vec t$, $\vec u$) or total ($\vec s$) momenta as
\begin{align}
t \equiv (t_0,\h \vec{t}) & = (\omega_3-\omega_2\hh, \h \vec{p}_3-\vec{p}_2) \,, \label{convt} \\[5pt]
u \equiv(u_0,\h \vec{u}) & = (\omega_1-\omega_3\hh,\h \vec{p}_1-\vec{p}_3) \,, \label{convu} \\[5pt]
s \equiv(s_0,\h \vec{s}) & = (\omega_1+\omega_2\hh,\h \vec{p}_1+\vec{p}_2) \,. \label{convs}
\end{align}
We introduce the following vertices, which differ from the original vertices merely in a relabeling of their arguments (for other conventions, see Refs.~\onlinecite{husemann, julian, Wang12}):
\begin{align}
 \varGamma_{k_1 k_2}(t) & := \varGamma(k_1, \h k_2 - t, \h k_2) \,, \label{gammat} \\[5pt]
 \varGamma_{k_1 k_2}(u) & := \varGamma(k_1, \h k_2 - u, \h k_1 - u) \,, \label{gammav} \\[5pt]
 \varGamma_{k_1 k_2}(s) & := \varGamma(k_1, \h s - k_1, \h s - k_2) \,. \label{gammas}
\end{align}
Note that the three functions on the left-hand side are actually different, although in our notation they are distinguished only by their respective argument ($t$, $u$, or~$s$). With these definitions, the parquet equations can be rewritten compactly as follows (for deriving the last equation, one has to use that $\varGamma^{\mathrm{pp}}$ is antisymmetric with respect to its first two arguments, see the appendix:
\begin{widetext}
\begin{align}
 \big[\varPhi^{\mathrm{ph,d}}_{\sigma_1 \sigma_2 \sigma_3 \sigma_4} \hh \big]_{k_1 k_2}(t) & = -\frac{k_{\mathrm B} T}{N} \sum_{k, \h \sigma, \h \sigma'} \big[F_{\sigma_1 \sigma' \sigma \sigma_4} \hh \big]_{k_1 k}(t) \,\h G(k) \, G(k - t) \, \big[\varGamma^{\mathrm{ph,d}}_{\sigma \sigma_2 \sigma_3 \sigma'}\big]_{k k_2}\mh(t) \,, \label{eq:parquet1_short} \\[5pt]
 \big[\varPhi^{\mathrm{ph,c}}_{\sigma_1 \sigma_2 \sigma_3 \sigma_4} \hh \big]_{k_1 k_2}(u) & = \frac{k_{\mathrm B} T}{N} \sum_{k, \h \sigma, \h \sigma'} \big[F_{\sigma_1 \sigma' \sigma_3 \sigma} \big]_{k_1 k}(u) \,\h G(k) \, G(k - u) \, \big[\varGamma^{\mathrm{ph,c}}_{\sigma \sigma_2 \sigma' \sigma_4}\big]_{k \hh k_2}\mh(u) \,, \label{eq:parquet2_short} \\[3pt]
 \big[\varPhi^{\mathrm{pp}}_{\sigma_1 \sigma_2 \sigma_3 \sigma_4}\big]_{k_1 k_2}(s) & = \frac 1 2 \h \frac{k_{\mathrm B} T}{N} \sum_{k, \h \sigma, \h \sigma'} \big[F_{\sigma_1 \sigma_2 \sigma \sigma'}]_{k_1 k}(s) \,\h G(k) \, G(s- k) \, \big[\varGamma^{\mathrm{pp}}_{\sigma'\sigma\sigma_3\sigma_4}\big]_{k \hh k_2}\mh(s) \,. \label{eq:parquet3_short}
\end{align}
\end{widetext}
We generally expect that the three reducible vertices $\varPhi^{\mathrm{ph,d}}$, $\varPhi^{\mathrm{ph,c}}$, $\varPhi^{\mathrm{pp}}$,
which together solve the parquet equations, have a strong dependence on their respective main transfer or total momentum (see e.g.~Refs.~\onlinecite{husemann, husemann12, eberlein10, maier2013, wang, julian}).
On the other hand, the dependencies on the remaining two fermionic momenta are expected to be rather weak. 
Therefore, we assume that these dependencies can be described to a sufficient accuracy by using only few smooth basis functions. 
Thus, we consider a set of {\itshape form factors,} i.e., functions in the first Brillouin zone (BZ) denoted by
\begin{equation} \label{eq:basis}
\big\{ f_{\vec{\ell}}(\vec{k})\h;\, \vec{k}\in 1^{\text {st}}\,\text{BZ},\, \vec{\ell}\in \mathbb{Z} \times \mathbb{Z} \hh \big\} \,,
\end{equation}
which are assumed to be pairwise orthonormal and complete in the sense that
\begin{align}
\frac 1 N \sum_{\vec{k}}f_{\vec{\ell}}(\vec{k}) \h f_{\vec{\ell}'}^*(\vec{k}) & = \delta_{\vec{\ell} \vec{\ell}'}
\label{eq:complete} \,, \\[2pt]
\sum_{\vec{\ell}}f_{\vec{\ell}}(\vec{k}) \h f_{\vec{\ell}}^*(\vec{k}') & = N \h \delta_{\vec{k} \vec{k}'}.
\label{eq:ortho}
\end{align}
Later, the index $\vec \ell$ will label the sites of the two-dimensional square lattice (see Eq.~\eqref{eq:form}). For technical reasons, we further introduce the multi-indices $\ell = (\ell_0, \h \vec{\ell})$ and the ``frequency-dependent'' form factors
\begin{equation} \label{form_factors}
 f_\ell(k) := \delta_{\ell_0, \hh k_0} \, f_{\vec \ell}(\vec k) \,,
\end{equation}
which are orthonormal and complete in the sense that
\begin{align}
 \frac 1 N \sum_k f_{\ell}(k) \h f_{\ell'}^*(k) & = \delta_{\ell \ell'} \,, \label{eq:complete_freq} \\
 \sum_{\ell} f_{\ell}(k) \h f_{\ell}^*(k') & = N \h \delta_{k k'} \,. \label{eq:ortho_freq}
\end{align}
In fact, the following considerations would remain valid for more general, truly frequency-dependent form factors fulfilling Eqs.~\eqref{eq:complete_freq}--\eqref{eq:ortho_freq}.
Finally, we define the {\itshape projections} of an arbitrary vertex $\varGamma$ onto the direct particle-hole, crossed particle-hole and particle-particle channel as follows (cf.~Ref.~\onlinecite[Eqs.~(18)-(20)]{julian}):
\begin{align}
 \hat D[\varGamma]_{\ell_1 \ell_2}(t) & = \frac 1 {N^2} \sum_{k_1, \h k_2} \varGamma_{k_1 k_2}(t) \, f_{\ell_1}(k_1) \, f_{\ell_2}^*(k_2) \,, \label{eq:C_short} \\
 \hat C[\varGamma]_{\ell_1 \ell_2}(u) & = \frac 1 {N^2} \sum_{k_1, \h k_2} \varGamma_{k_1 k_2}(u) \, f_{\ell_1}(k_1) \, f_{\ell_2}^*(k_2) \,, \label{eq:D_short} \\
 \hat P[\varGamma]_{\ell_1 \ell_2}(s) & = \frac 1 {N^2} \sum_{k_1, \h k_2} \varGamma_{k_1 k_2}(s) \, f_{\ell_1}(k_1) \, f_{\ell_2}^*(k_2) \,. \label{eq:P_short}
\end{align}
By combining these equations with Eqs.~\eqref{gammat}--\eqref{gammas}, we obtain the following formulae, by which the vertex $\varGamma$ can be reconstructed from its respective projections (these equations will be used in Sct.~\ref{sec:cross} to derive the ``cross projections'' between different channels):
\begin{widetext}
\begin{align}
 \varGamma(p_1, p_2, p_3) & = \sum_{\ell_1, \h \ell_2} \hat D[\varGamma]_{\ell_1 \ell_2}(p_3 - p_2) \, f_{\ell_1}^*(p_1) \, f_{\ell_2}(p_3) \,, \label{eq:C_short_inverse} \\[5pt]
 \varGamma(p_1, p_2, p_3) & = \sum_{\ell_1, \h \ell_2} \hat C[\varGamma]_{\ell_1 \ell_2}(p_1 - p_3) \, f_{\ell_1}^*(p_1) \, f_{\ell_2}(p_1 + p_2 - p_3) \,, \label{eq:D_short_inverse} \\[5pt]
 \varGamma(p_1, p_2, p_3) & = \sum_{\ell_1, \h \ell_2} \hat P[\varGamma]_{\ell_1 \ell_2}(p_1 + p_2) \, f_{\ell_1}^*(p_1) \, f_{\ell_2}(p_1 + p_2 - p_3) \,. \label{eq:P_short_inverse} 
 \end{align}
\end{widetext}
Using these relations, 
we will now derive parquet-type equations for the reducible vertices 
$\varPhi^{\mathrm{ph,d}}$, $\varPhi^{\mathrm{ph,c}}$, $\varPhi^{\mathrm{pp}}$, where each of these vertices is
projected onto its respective main momentum. 
For this purpose, we apply the mappings \eqref{eq:C_short}--\eqref{eq:P_short} to both sides of Eqs.~\eqref{eq:parquet1_short}--\eqref{eq:parquet3_short}
and insert two {\itshape partitions of unity} of the form-factor basis (see Eq.~\eqref{eq:ortho_freq}) on both sides of the fermion loops.
This procedure is analogous to the derivation of flow equations
in the truncated-unity functional renormalization group (TUfRG) scheme \cite{julian,David17, Schober17}.
We thus arrive at the following self-consistent equations, which we call the {\itshape truncated-unity (TU) parquet equations:}
\begin{widetext}
\begin{align}
\hat{D}\big[\varPhi^{\mathrm{ph,d}}_{\sigma_1 \sigma_2 \sigma_3 \sigma_4}\big]_{\ell_1 \ell_2} \mh(t) &=
- \sum_{\ell, \h \ell'\mh, \h \sigma, \h \sigma'} \hat{D}\big[F_{\sigma_1 \sigma' \sigma \sigma_4}\big]_{\ell_1\ell}\h(t) \, \h L^{\mathrm{ph}}_{\ell \ell'}(t) \,\h \hat{D}\big[\varGamma^{\mathrm{ph,d}}_{\sigma \sigma_2 \sigma_3 \sigma'}\big]_{\ell' \ell_2} \mh (t) \,,
\label{eq:projected1}
\\[5pt]
\hat{C}\big[\varPhi^{\mathrm{ph,c}}_{\sigma_1 \sigma_2 \sigma_3 \sigma_4}\big]_{\ell_1 \ell_2} \mh(u) &=
\sum_{\ell, \h \ell'\mh, \h \sigma, \h \sigma'} \hat{C}\big[F_{\sigma_1 \sigma' \sigma_3 \sigma}\big]_{\ell_1\ell}\h(u) \, \h L^{\mathrm{ph}}_{\ell \ell'}(u) \,\h \hat{C}\big[\varGamma^{\mathrm{ph,c}}_{\sigma \sigma_2 \sigma' \sigma_4}\big]_{\ell' \ell_2} \mh (u) \,,
\label{eq:projected2}\\[5pt]
\hat{P}\big[\varPhi^{\mathrm{pp}}_{\sigma_1 \sigma_2 \sigma_3 \sigma_4}\big]_{\ell_1 \ell_2}(s) &= \frac{1}{2} \, \sum_{\ell, \h \ell'\mh, \h \sigma, \h \sigma'} \hat P\big[F_{\sigma_1 \sigma_2 \sigma \sigma'}\big]_{\ell_1 \ell}\h(s) \,\h L^{\mathrm{pp}}_{\ell \ell'}(s) \,\h \hat P\big[\varGamma^{\mathrm{pp}}_{\sigma'\sigma\sigma_3 \sigma_4}\big]_{\ell' \ell_2}\mh(s) \,,
\label{eq:projected3}
\end{align}
where the {\itshape particle-hole loop} $L^{\mathrm{ph}}$ and the {\itshape particle-particle loop} $L^{\mathrm{pp}}$ are given by
\begin{align}
 L^{\mathrm{ph}}_{\ell \ell'}(t) & = \frac{k_{\mathrm B} T} N \sum_k G(k) \, G(k - t) \, f_{\ell}(k) \, f_{\ell'}^*(k) \,, \label{loop_ph} \\[5pt]
 L^{\mathrm{pp}}_{\ell \ell'}(s) & = \frac{k_{\mathrm B} T} N \sum_k G (k) \, G(s - k) \, f_{\ell}(k) \, f_{\ell'}^*(k) \,. \label{loop_pp}
\end{align}
\end{widetext}
One main feature of Eqs.~\eqref{eq:projected1}--\eqref{eq:projected3} is that they involve only matrix multiplications with respect to the internal summation indices $\ell$ and $\ell'$ of the form-factor basis (provided one uses a countable set of form factors).
This structure of the TU parquet equations is particularly advantageous for the numerical parallelization because
it allows for an independent evaluation for different values of $t$, $u$, and $s$, and thus for a distribution of the vertices over several compute nodes. We remark, however, that internode communication is still needed when invoking Eq.~\eqref{eq:Fsum}. Furthermore, the above form of the parquet equations still requires the calculation of cross projections between different channels as we will explain in the following.

\section{Cross projections} \label{sec:cross}

The standard procedure for solving the self-consistent parquet equations is an iteration scheme, which takes the bare interaction as the initial vertex and in each step evaluates the parquet equations once to recalculate the vertices. To explain this in more detail for the TU parquet equations \eqref{eq:projected1}--\eqref{eq:projected3}, let us assume that in one iteration step we have calculated the projected vertices
$\hat D[\varPhi^{\rm ph,d}]$, $\hat C[\varPhi^{\mathrm{ph,c}}]$, and $\hat P[\varPhi^{\mathrm{pp}}]$.
Then, in the next step, we want to employ Eqs.~\eqref{eq:projected1}--\eqref{eq:projected3} to recalculate these vertices. Consider, as an example, the projection $\hat{D}[F]$ of the total vertex, which appears on the right-hand side of Eq.~\eqref{eq:projected1}. This can be split into the initial interaction and the three channels (see Eq.~\eqref{eq:Fsum}), i.e.,
\begin{equation} \label{eq:CF1}
\begin{aligned}
\hat{D}[F]_{\ell_1 \ell}(t) & = 
\hat{D}[F^0]_{\ell_1 \ell} + \hat{D}[\varPhi^{\mathrm{ph,d}}]_{\ell_1 \ell}(t) \\[5pt]
& \quad \, +
\hat{D}[\varPhi^{\mathrm{ph,c}}]_{\ell_1 \ell}(t) +
\hat{D}[\varPhi^{\mathrm{pp}}]_{\ell_1 \ell}(t)\,.
\end{aligned}
\end{equation}
\smallskip
Similarly, $\hat D[\varGamma^{\mathrm{ph,d}}]$ can be calculated from Eq.~\eqref{eq:Fsplit} as
\begin{equation}
\hat{D}[\varGamma^{\mathrm{ph,d}}]_{\ell' \ell_2}(t) = \hat{D}[F]_{\ell' \ell_2}(t) - \hat{D}[\varPhi^{\mathrm{ph,d}}]_{\ell' \ell_2}(t) \,.
\label{eq:gam}
\end{equation}
Now, the first term in Eq.~\eqref{eq:CF1}, i.e., the projection of the initial interaction, is known explicitly: combining Eqs.~\eqref{ini_int} and \eqref{eq:C_short} gives
\begin{align}
 \hat D[F^0]_{\ell_1 \ell} & = \frac 1 {N^2} \sum_{k, \h k'} f_{\ell_1}(k) \h f_{\ell_2}^*(k') \, U \\[3pt]
 & \equiv \h \langle f_{\ell_1} \rangle \h \langle f_\ell^* \rangle \, U \,, \label{D_ini}
\end{align}
where we have omitted the spin dependencies. The second term in Eq.~\eqref{eq:CF1}, namely
$\hat{D}[\varPhi^{\mathrm{ph,d}}]$, is directly available from the previous iteration step. By contrast, for calculating the projections
$\hat{D}[\varPhi^{\mathrm{ph,c}}]$ and $\hat{D}[\varPhi^{\mathrm{pp}}]$ from the previously obtained $\hat{C}[\varPhi^{\mathrm{ph,c}}]$ and $\hat P[\varPhi^{\mathrm{pp}}]$, it is necessary to invert the projections $\hat{C}$ and $\hat{P}$. Hence, we can calculate $\hat D[F]$ by means of the formal identity

\begin{equation} 
\begin{aligned}
\hat{D}[F]_{\ell_1 \ell}(t) & = 
\hat{D}[F^0]_{\ell_1 \ell} + \hat{D}[\varPhi^{\mathrm{ph,d}}]_{\ell_1 \ell}(t) \\[5pt]
& \quad \, + \hat{D}[\hat{C}^{-1}[\hat{C}[\varPhi^{\mathrm{ph,c}}]]]_{\ell_1 \ell}(t) \\[5pt]
& \quad \, + \hat{D}[\hat{P}^{-1}[\hat{P}[\varPhi^{\mathrm{pp}}]]]_{\ell_1 \ell}(t) \,.\label{eq:CF2}
\end{aligned}
\end{equation}
\smallskip

To calculate the ``cross projections'' between the different channels, we may use the definitions \eqref{eq:C_short}--\eqref{eq:P_short} as well as Eqs.~\eqref{eq:C_short_inverse}--\eqref{eq:P_short_inverse}. After some algebra, we thus obtain
\begin{widetext}
\begin{equation} \label{eq:CF3}
\begin{aligned}
\hat{D}[F]_{\ell_1 \ell}(t) & =
\hat{D}[F^0]_{\ell_1 \ell} + \hat{D}[\varPhi^{\mathrm{ph,d}}]_{\ell_1 \ell}(t) \\[5pt]
& \quad \, + \frac 1 {N^2} \sum_{k'} \sum_{\ell_3, \h \ell_4} \hat C[\varPhi^{\mathrm{ph,c}}]_{\ell_3 \ell_4}(k') \, \sum_{k_1} f_{\ell_3}^*(k_1) \, f_{\ell_4}(k_1 - t) \, f_{\ell_1}(k_1) \, f_{\ell}^*(k_1 - k') \\[3pt]
& \quad \, + \frac 1 {N^2} \sum_{k'} \sum_{\ell_3, \h \ell_4} \hat P[\varPhi^{\mathrm{pp}}]_{\ell_3 \ell_4}(k') \, \sum_{k_1} f_{\ell_3}^*(k_1) \, f_{\ell_4}(k_1 - t) \, f_{\ell_1}(k_1) \, f_{\ell}^*(k' - k_1 + t) \,.
\end{aligned}
\end{equation}
\end{widetext}
Similarly, one can derive the cross projections between any two other channels, and this then allows one to iteratively solve the TU parquet equations.

We go on to describe some approximations which further simplify the iterative solution. First, we neglect the frequency dependencies of all vertices. The remaining frequency sums in the fermion loops \eqref{loop_ph}--\eqref{loop_pp} can be performed analytically \cite{Mahan}, giving 
\begin{widetext}
\begin{align}
L^{\mathrm{ph}}_{\vec \ell \vec \ell'}(\vec t) & := \sum_{\ell_0, \h \ell'_0} L^{\mathrm{ph}}_{\ell \ell'}(\vec t, t_0 = 0) 
\h = \h \frac 1 N \sum_{\vec k} \frac{n_{\mathrm F}(\varepsilon(\vec k)) - n_{\mathrm F}(\varepsilon(\vec k - \vec t))}{\varepsilon(\vec k) - \varepsilon(\vec k - \vec t)} \, f_{\vec \ell}(\vec k) \, f_{\vec \ell'}^*(\vec k) \,, \label{ph_l} \\[5pt]
L^{\mathrm{pp}}_{\vec \ell \vec \ell'}(\vec s) & := \sum_{\ell_0, \h \ell'_0} L^{\mathrm{pp}}_{\ell \ell'}(\vec s, s_0 = 0) \h = \h \frac 1 N \sum_{\vec k} \frac{1 - n_{\mathrm F}(\varepsilon(\vec k)) - n_{\mathrm F}(\varepsilon(\vec s - \vec k))}{\varepsilon(\vec k) + \varepsilon(\vec s - \vec k)} \, f_{\vec \ell}(\vec k) \, f_{\vec \ell'}^*(\vec k) \,. \label{pp_l}
\end{align}
\end{widetext}
Here, $n_{\mathrm F}(\varepsilon) = (1 + \exp(\beta\varepsilon))^{-1}$ denotes the Fermi distribution function, which depends on the inverse temperature $\beta = 1/k_{\mathrm B} T$.
Next, as has already been seen in (Ref.~\onlinecite{wang}; see also Sct.~3.3 of Ref.~\onlinecite{Schober17}), a further simplification can be achieved by using complex exponentials as form factors, i.e.,
\begin{equation} \label{eq:form}
f_{\vec{\ell}}(\vec k) =\mathrm{e}^{-\mathrm{i} a \hh \vec{\ell} \hh \cdot \hh \vec k} \,,
\end{equation}
where $a$ denotes the lattice spacing. This set of functions naturally fullfills the requirements \eqref{eq:complete}--\eqref{eq:ortho}. Then, Eq.~\eqref{D_ini} simplifies to
\begin{equation}
\hat D[F^0]_{\ell_1 \ell} = U \, \delta_{\vec \ell_1 \mhh, \hh \vec 0} \, \delta_{\vec \ell, \hh \vec 0} \,.
\end{equation}
Furthermore, by introducing Fourier-transformed projections $\widetilde D$, $\widetilde C$, $\widetilde P$, such that
\begin{equation} \label{eq:FT1}
 \hat D[F]_{\vec \ell_1 \vec \ell}(\vec t) = \sum_{\vec \ell'} \mathrm e^{-\mathrm i a \hh \vec \ell' \mhh \cdot \hh \vec t} \h \widetilde D[F]_{\vec \ell_1 \vec \ell}\h(\vec \ell') \,,
\end{equation}
or conversely,
\begin{equation}
 \widetilde D[F]_{\vec \ell_1 \vec \ell}(\vec \ell') = \frac 1 N \sum_{\vec t} \mathrm e^{\mathrm i a \hh \vec \ell' \mhh \cdot \hh \vec t} \h \hat D[F]_{\vec \ell_1 \vec \ell}\h(\vec t) \,, \label{eq:FT2}
\end{equation}
we can transform Eq.~\eqref{eq:CF3} into
\begin{align} \label{eq:CF4}
 & \hat D[F]_{\vec \ell_1 \vec \ell}(\vec t) = U \, \delta_{\vec \ell_1 \mhh, \hh \vec 0} \, \delta_{\vec \ell, \hh \vec 0} + \hat D[\varPhi^{\mathrm{ph,d}}]_{\vec \ell_1 \vec \ell}(\vec t) \\[5pt] \nonumber
 & + \h \sum_{\vec \ell'} \mathrm{e}^{\mathrm i a \hh \vec \ell' \mhh \cdot \hh \vec t} \,\hh \widetilde C[\varPhi^{\mathrm{ph,c}}]_{\vec \ell' + \vec \ell_1 - \vec \ell, \, \vec \ell'}\hh(-\vec \ell) \\ \nonumber
 & + \h \sum_{\vec \ell'} \mathrm{e}^{\mathrm i a \hh \vec \ell' \mhh \cdot \hh \vec t} \,\hh \widetilde P[\varPhi^{\mathrm{pp}}]_{\vec \ell' + \vec \ell_1, \, \vec \ell' - \vec \ell}\h(\vec \ell) \,.
\end{align}
We remark that instead of simple plane-wave functions given by Eq.~\eqref{eq:form}, one could also use form factors that explicitly respect the symmetry of the lattice, so-called lattice harmonics.
Fewer of these are required to reach the same accuracy as with simple plane-wave functions. For an introduction to lattice harmonics, we refer the interested reader to Sct.~A3 of Ref.~\onlinecite{platt13}.

Finally, the main approximation which the channel projetions aim at is to keep only a finite number of form factors $f_{\vec \ell}(\vec k)$, such that
\begin{equation} \label{eq:cutoff}
 \vec{\ell}\in [-\ell_{\text{cut}}, \h \ell_{\text{cut}}]\times[-\ell_{\text{cut}}, \h \ell_{\text{cut}}] \,,
 \end{equation}
with a cutoff parameter $\ell_{\text{cut}} \in \mathbb N$.
The total number of form factors is then $n_{\mathrm{cut}} \equiv 4 \h \ell_{\mathrm{cut}}^2$\hh.
To illustrate the implications of this approximation, we consider again Eq.~\eqref{eq:CF3}.
For each particular combination of arguments $\ell_1$, $\ell$, and $t$, the projected vertex $\hat{D}[F]_{\ell_1 \ell}(t)$ depends on $\hat{C}[\varPhi^{\mathrm{ph,c}}]_{\ell_3 \ell_4}(k')$ (and similarly on $\hat{P}[\varPhi^{\mathrm{pp}}]_{\ell_3,\ell_4}(k')$) at all possible arguments $\ell_3$, $\ell_4$, and $k'$. Similar considerations hold for the Fourier-transformed vertices as in Eq.~\eqref{eq:CF4}.
In other words, it is in principle necessary to know the projected vertices {\itshape at all possible arguments} in one iteration step
before one can recalculate these vertices {\itshape at any particular argument} in the next iteration step.
Therefore, when keeping only a limited set of basis functions as specified by Eq.~\eqref{eq:cutoff},
the vertices $\hat{C}[\varPhi^{\mathrm{ph,c}}]_{\ell_3 \ell_4}$ and $\hat{P}[\varPhi^{\mathrm{pp}}]_{\ell_3 \ell_4}$ (or their Fourier transforms)
from one iteration step contribute to $\hat{D}[F]_{\ell_1 \ell}$ in the next iteration step only approximately (in the sense that also their strong dependence on the main momentum is not accounted for exactly).
This in turn reflects the fact that the projections $\hat{D}$, $\hat{C}$, and $\hat{P}$ are actually not invertible if restricted to a limited set of form factors.
Nevertheless, the inverse mappings $\hat{C}^{-1}$ and $\hat{P}^{-1}$ as formally employed in Eq.~\eqref{eq:CF2} can be approximated by means of Eq.~\eqref{eq:CF3} or Eq.~\eqref{eq:CF4}, provided that a sufficiently large number of form factors is taken into account in these summations.
 
To summarize, it is not {\itshape a priori} clear why keeping only a limited number of form factors is a good approximation.
In particular, if the vertex has some sharp momentum structures in one channel (close to a phase transition at low temperatures, or caused by long-range initial interactions), then these sharp structures in one channel would in principle also lead to sharp structures in the other channels by means of the cross projections.
However, keeping only a few form factors would cause these sharp structures to be smeared out.
Notwithstanding this {\itshape caveat,} we expect that local contributions to the vertices are more important for determining critical temperatures.
Previous fRG studies (e.g.~Ref.~\onlinecite{julian}) have shown that for the $t-t'$ Hubbard Model, including the Fermi-surface studied here
\footnote{The plot in the cited paper does not explicitly cover half filling.
The authors of the present manuscript have, however, convinced themselves that the same conclusion also holds for half-filling.
This has been done together with the authors of Ref.~\onlinecite{julian} using the same code as in the cited work.}
, the inclusion of higher-order form factors does not change the critical scales considerably.

\section{Complexity and memory cost} \label{sec:numerical}

In this section, we briefly discuss the computational complexity and memory cost of our projection scheme
as compared to a direct implementation of the parquet equations.
We limit this discussion to the evaluation of the momentum dependencies of the vertices, since the other dependencies
(i.e., those on spin and frequency) are not affected by the projections (see Eqs.~\eqref{form_factors} and \eqref{eq:C_short}--\eqref{eq:P_short}). 

First, consider a direct evaluation of the parquet equations \eqref{eq:parquet1_full}--\eqref{eq:parquet3_full}.
For each combination of three ``external'' momentum arguments, a sum over one ``internal'' momentum has to be performed. Since each momentum ranges in the first BZ, the complexity of evaluating these equations is $\mathcal{O}(N^4)$,
where $N$ is the number of discrete Bloch momenta.
This can possibly be improved by using matrix multiplications, for which efficient algorithms exist,
but a lower bound is always $\mathcal{O}(N^3)$ corresponding to the memory cost.
Furthermore, since vertices of three momentum arguments have to be stored, the memory consumption scales like $\mathcal{O}(N^3)$.

Next, consider the channel-decomposed parquet equations \eqref{eq:projected1}-\eqref{eq:projected3}. For each argument of the vertices on the left-hand side, two sums over the form-factor basis have to be performed. Since the vertices have $N \times n_{\text{cut}}^{2}$ arguments, this scales like $\mathcal{O}(N \times n_{\text{cut}}^{4})$. In order to evaluate the parquet equations, one further needs the projections of $F$ onto each channel as well as the loop functions $L^{\mathrm{ph}}$ and $L^{\mathrm{pp}}$.
Note that only the cross projection of $F$ has to be computed, because the one of $\varGamma^{r}$ can simply be obtained via Eq.~\eqref{eq:gam}. For evaluating the cross projections we have presented two different schemes, namely Eq.~\eqref{eq:CF3} and Eq.~\eqref{eq:CF4}. In the first case, one has to perform one momentum sum and two form-factor sums  for each argument of the projected vertex, which scales like $\mathcal{O}(N^{2} \times n_{\text{cut}}^{4})$.
The second sum in Eq.~\eqref{eq:CF3} (over $k_1$) can be computed in advance and is therefore faster than the other calculations.
In the second case, i.e., when using Eq.~\eqref{eq:CF4}, only one sum over the form-factor basis has to be performed.
Again, this has to be done for each argument of the projected vertex, thus scaling like $\mathcal{O}(N \times n_{\text{cut}}^{3})$. Finally, one still needs to calculate the bubble functions via Eqs.~\eqref{loop_ph}--\eqref{loop_pp}. Here, one momentum sum is required for each argument of the bubbles, thus scaling like $\mathcal{O}(N^{2} \times n_{\text{cut}}^{2})$.

We conclude that when using Eq.~\eqref{eq:CF3}, the cross projections are actually the most expensive calculations such that the overall complexity is $\mathcal{O}(N^{2} \times n_{\text{cut}}^{4})$. On the other hand, when invoking Eq.~\eqref{eq:CF3}, the most expensive calculation is the evaluation of the bubble functions.
However, these bubble functions need to be calculated only once because they do not change during the iteration process (at least if no self-energy is calculated). Hence, in this second case the amortized costs are determined by the evaluation of Eqs.~\eqref{eq:projected1}-\eqref{eq:projected3}, and therefore the overall amortized costs scale like $\mathcal{O}(N \times n_{\text{cut}}^{4})$. In particular, the complexity then scales only linearly with the momentum resolution.

Concerning the memory costs, all vertices have to be stored as functions of their main momentum and two basis-function indices, which implies a scaling like $\mathcal{O}(N \times n_{\text{cut}}^2)$.
This does not change when employing Eq.~\eqref{eq:CF4}: for this scheme, the Fourier-transformed vertices would have to be stored in addition to the original vertices. The Fourier-transformed vertices are not larger than the original vertices, but if the vertices are already highly memory-consuming, the additional storage space required might still not be negligible.
Thus, if memory is the main constraint, one can instead employ Eq.~\eqref{eq:CF3} at the cost of a higher complexity.
In summary, if only a constant set of basis functions is kept, the memory consumption scales only linearly with the momentum resolution $N$.

Finally, we remark that previous implementations of the parquet equations \cite{held_new} have shown that the ``bottle neck'' for evaluating them is in fact the memory consumption. In particular, when performing computations on heavily parallel machines, the internode communications usually limit the system sizes that can be treated. Thus, reducing the memory consumption to linear scaling in $N$ as in the present TU parquet approach may indeed represent a major step forward.

\section{Numerical implementation} \label{sec:numerical_implementation}

In order to check the validity of our method, we have applied the TU parquet equations to the Hubbard model on the square lattice given by the Hamiltonian
\begin{equation} \label{Hub_Ham}
 H = -t \sum_{\langle i, \hh j \rangle, \h \sigma} a_{i, \sigma}^\dagger \h a_{j, \sigma} + U \h \sum_{i} n_{i, \uparrow} \h n_{i, \downarrow} \,.
\end{equation}
Here, $a_{i, \sigma}$ and $a_{i, \sigma}^\dagger$ denote the annihilation and creation operators at site $i$ with spin $\sigma \in \{\uparrow, \downarrow\}$, and
$n_{i, \sigma} = a_{i, \sigma}^{\dagger} \h a_{i, \sigma}$ the number operator at site $i$. 
The sum in the first term in Eq.~\eqref{Hub_Ham} is only over nearest-neighbor sites, while the second term describes an onsite interaction. By diagonalizing the Hamiltonian, one obtains the energy dispersion
\begin{equation}
 \varepsilon(\vec k) = 2 \hh t \h (\cos(k_x \hh a) + \cos(k_y \hh a) ) \,. \medskip
\end{equation}
The Hubbard model has been studied extensively in the literature (for some references, see e.g.~Refs.~\onlinecite{maierDCA,gull,schaefer}) as it is expected to be relevant for the high-temperature superconducting cuprates \cite{scalapino}. Therefore, this model can serve as a testing case for any newly developed quantum many-body method.

Our numerical solution of the TU parquet equations uses the approximations mentioned in Sct.~\ref{sec:cross}, hence we neglect all frequency dependencies of the vertices, setting $t_0 = u_0 = s_0 = 0$.
As form factors we choose complex exponentials, which have already been given in Eq.~\eqref{eq:form}.
Furthermore, we neglect the self-energy and thereby replace the full Green function with the bare Green function: $G = G_0$.
Finally, as our aim here is a consistency check, we limit the discussion to the \mbox{$\vec{\ell} = \vec 0$} contributions (thus considering only a single basis function) in the TU parquet equations.
It is known, however, that the critical scales in TUfRG for the same model at half filling change only insignificantly when higher form factors $\vec \ell \not = \vec 0$ are taken into account \cite{julian}.
Of course, this changes when one moves away from half filling and non-local pairing becomes important. In any case, within the ``onsite'' $\vec{\ell} = \vec 0$ approximation, the evaluation of the parquet equations becomes particularly simple because then Eq.~\eqref{eq:CF4} reduces to
\begin{equation}
\begin{aligned} 
\hat D[F]_{\vec 0 \vec 0}(\vec t) & = U + \hat D[\varPhi^{\mathrm{ph,d}}]_{\vec 0 \vec 0}(\vec t) \nonumber \\[5pt] \label{eq:simpler}
& \quad \, + \big \langle \hat C[\varPhi^{\mathrm{ph,c}}]_{\vec 0 \vec 0} \big \rangle + \big\langle \hat P[\varPhi^{\mathrm{pp}}]_{\vec 0 \vec 0} \big\rangle \,.
\end{aligned}
\end{equation}
Here, $\langle . \rangle$ denotes the mean of a vertex with respect to its main momentum argument (which coincides with the Fourier-transformed vertex evaluated at zero), i.e.,
\begin{equation} \label{eq:av}
\langle \varGamma \rangle := \frac 1 N \sum_{\vec t} \varGamma(\vec t) \,.
\end{equation}
Thus, the projection operation $\hat C[\hat D^{-1}[\hat D[\, . \,]]$ simply reduces to taking the mean of $\hat D[\, . \,]$ with respect to its main momentum argument, and the same applies to the other cross projections.

Before writing out the parquet equations within the above approximations, let us briefly discuss the spin dependence of the vertices.
In fact, since the Hubbard model has an SU(2)-symmetry (see e.g.~Ref.~\onlinecite{SH00}), it is possible to eliminate all spin indices and thereby to simplify the parquet equations.
We only state the results here and refer the interested reader to the appendix for a detailed derivation.
Generally, SU(2)-symmetry implies the following spin dependence of the vertices as a consequence of the so-called crossing relations \cite{bickers}:
\begin{widetext}
\begin{align}
\varPhi^{\mathrm{ph,d}}_{\sigma_1 \sigma_2 \sigma_3 \sigma_4}(k_1,k_2,k_3,k_4) & = -V^{\mathrm{ph,d}}(k_1,k_2,k_3,k_4) \, \delta_{\sigma_1 \sigma_4}\h\delta_{\sigma_2 \sigma_3} + V^{\mathrm{ph,c}}(k_1,k_2,k_4,k_3) \, \delta_{\sigma_1 \sigma_3}\h\delta_{\sigma_2 \sigma_4} \,, \label{eq_cross_1} \\[5pt]
\varPhi^{\mathrm{ph,c}}_{\sigma_1 \sigma_2 \sigma_3 \sigma_4}(k_1,k_2,k_3,k_4) & = -V^{\mathrm{ph,c}}(k_1,k_2,k_3,k_4) \, \delta_{\sigma_1 \sigma_4}\h\delta_{\sigma_2 \sigma_3} + V^{\mathrm{ph,d}}(k_1,k_2,k_4,k_3) \, \delta_{\sigma_1 \sigma_3}\h\delta_{\sigma_2 \sigma_4} \,, \label{eq_cross_2} \\[5pt]
\varPhi^{\mathrm{pp}}_{\sigma_1 \sigma_2 \sigma_3 \sigma_4}(k_1,k_2,k_3,k_4) & = -V^{\mathrm{pp}}(k_1,k_2,k_3,k_4) \, \delta_{\sigma_1 \sigma_4}\h\delta_{\sigma_2 \sigma_3} + V^{\mathrm{pp}}(k_1,k_2,k_4,k_3) \, \delta_{\sigma_1 \sigma_3}\h\delta_{\sigma_2 \sigma_4} \,. \label{eq_cross_3}
\end{align}
\end{widetext}
Conversely, the spin-independent $V$-functions can be obtained from the spin-dependent vertices by evaluating the latter at particular spin combinations, i.e.,\begin{equation}
V^{\mathrm{ph,d}}(k_1, k_2, k_3, k_4) = -\varPhi^{\mathrm{ph,d}}_{\uparrow \downarrow \downarrow \uparrow }(k_1, k_2, k_3, k_4) \,,
\end{equation}
and similarly for the other vertices. 
Furthermore, by evaluating the parquet equations \eqref{eq:parquet1_full}--\eqref{eq:parquet3_full} at spin arguments $(\sigma_1 \sigma_2 \sigma_3 \sigma_4) = (\uparrow\downarrow\downarrow\uparrow)$ and using Eqs.~\eqref{eq_cross_1}--\eqref{eq_cross_3}, one can derive the corresponding parquet equations for the $V$-functions. These in turn can be transformed into a channel-decomposed version as shown in the appendix.

In summary, we have implemented approximate TU parquet equations for the $\vec 0 \vec 0$-components of the channel-projected $V$-functions, which we abbreviate as
\begin{align}
 D(\vec t) & := \hat D[V^{\mathrm{ph, d}}]_{\vec 0 \vec 0}(\vec t) \,, \label{final_D} \\[3pt]
 C(\vec u) & := \hat C[V^{\mathrm{ph, c}}]_{\vec 0 \vec 0}(\vec u) \,, \label{final_C} \\[3pt]
 P(\vec s) & := \hat P[V^{\mathrm{pp}}]_{\vec 0 \vec 0}(\vec s) \,. \label{final_P}
\end{align}
The equations which we have implemented read as follows (where averages are defined as in Eq.~\eqref{eq:av}, and where the $\vec 0 \vec 0$-components of the fermion loops can be read off from Eqs.~\eqref{ph_l}--\eqref{pp_l}):
\begin{widetext}
\begin{align} \label{eq:PA_projected_1}
 D(\vec t) & = 2 \, \big( \hh U + D(\vec t) + \langle C \rangle + \langle P \rangle \hh \big) \, L^{\mathrm{ph}}_{\vec 0 \vec 0}(\vec t) \, \big( \hh U + \langle C \rangle + \langle P \rangle \hh \big) \\[3pt] \nonumber
 & \quad \, - \big( \hh U + D(\vec t) + \langle C \rangle + \langle P \rangle \hh \big) \, L^{\mathrm{ph}}_{\vec 0 \vec 0}(\vec t) \, \big( \hh U + \langle D \rangle + \langle P \rangle \hh \big) \\[3pt] \nonumber
 & \quad \, - \big( \hh U + \langle D \rangle + C(\vec t) + \langle P \rangle \hh \big) \, L^{\mathrm{ph}}_{\vec 0 \vec 0}(\vec t) \, \big( \hh U + \langle C \rangle + \langle P \rangle \hh \big) \,, 
 \\[8pt]
C(\vec u) & = - \, \big( \hh U + \langle D \rangle + C(\vec u) + \langle P \rangle \hh \big) \, L^{\mathrm{ph}}_{\vec 0 \vec 0}(\vec u) \, \big( \hh U + \langle D \rangle + \langle P \rangle \hh \big) \,, \label{eq:PA_projected_2} \\[8pt]
P(\vec s) & = - \, \big( \hh U + \langle D \rangle + \langle C \rangle + P(\vec s) \hh \big) \, L^{\mathrm{pp}}_{\vec 0 \vec 0}(\vec s) \, \big( \hh U + \langle D \rangle + \langle C \rangle \hh \big) \,. \label{eq:PA_projected_3}
 \end{align}
\end{widetext}
Importantly, since the fermion loops as well as all mean values of the vertices can be computed in advance, the numerical evaluation of these equations scales only linearly with the momentum resolution.

\section{Results for the Hubbard model} \label{sec:results}

\begin{figure}[t]
\begin{subfigure}{0.5 \textwidth}
\centering
\includegraphics[scale=1.0]{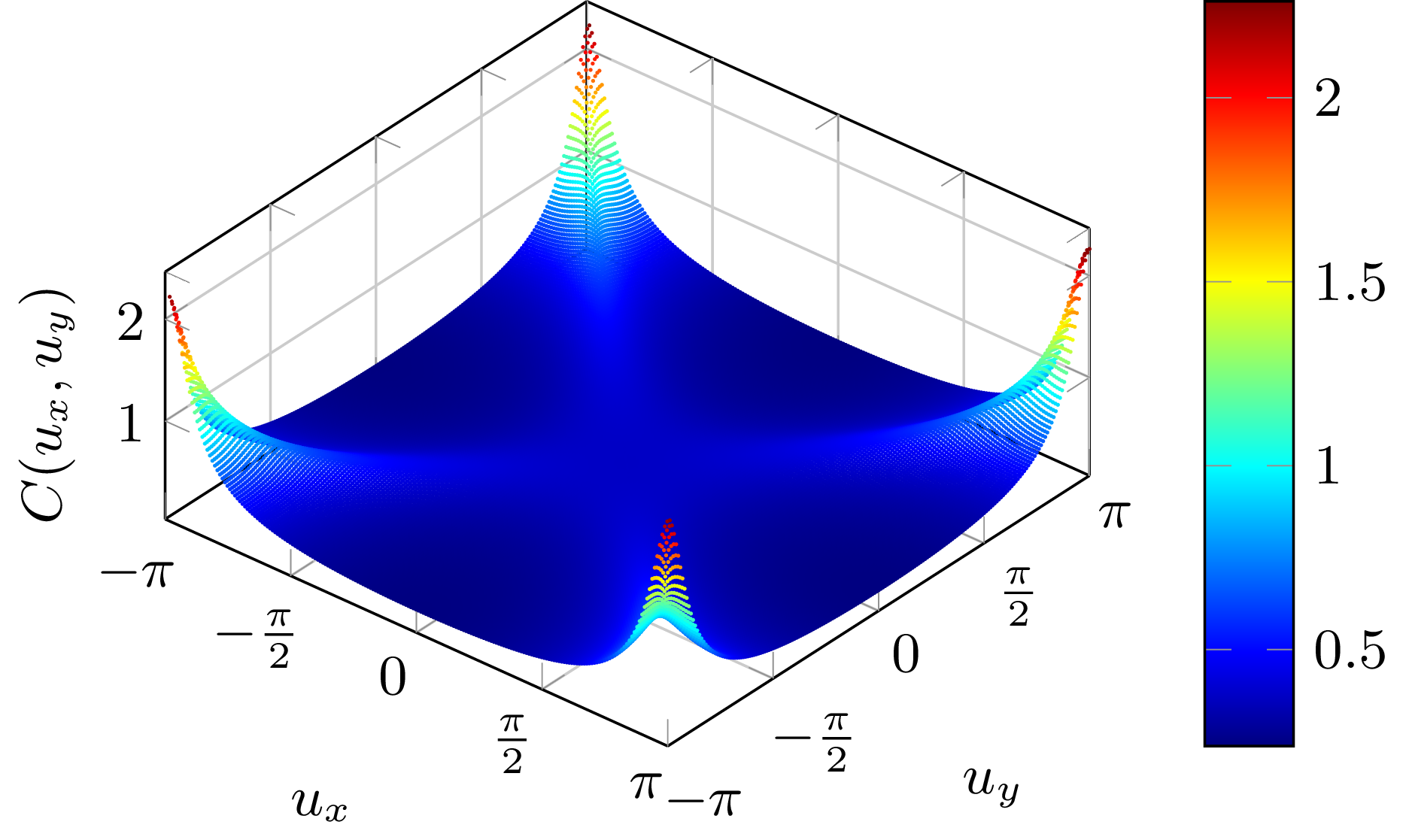}
\caption{Crossed particle-hole vertex $C(\vec u)$. \label{fig:ph}}
\end{subfigure}
\begin{subfigure}{0.5 \textwidth}
\includegraphics[scale=1.0]{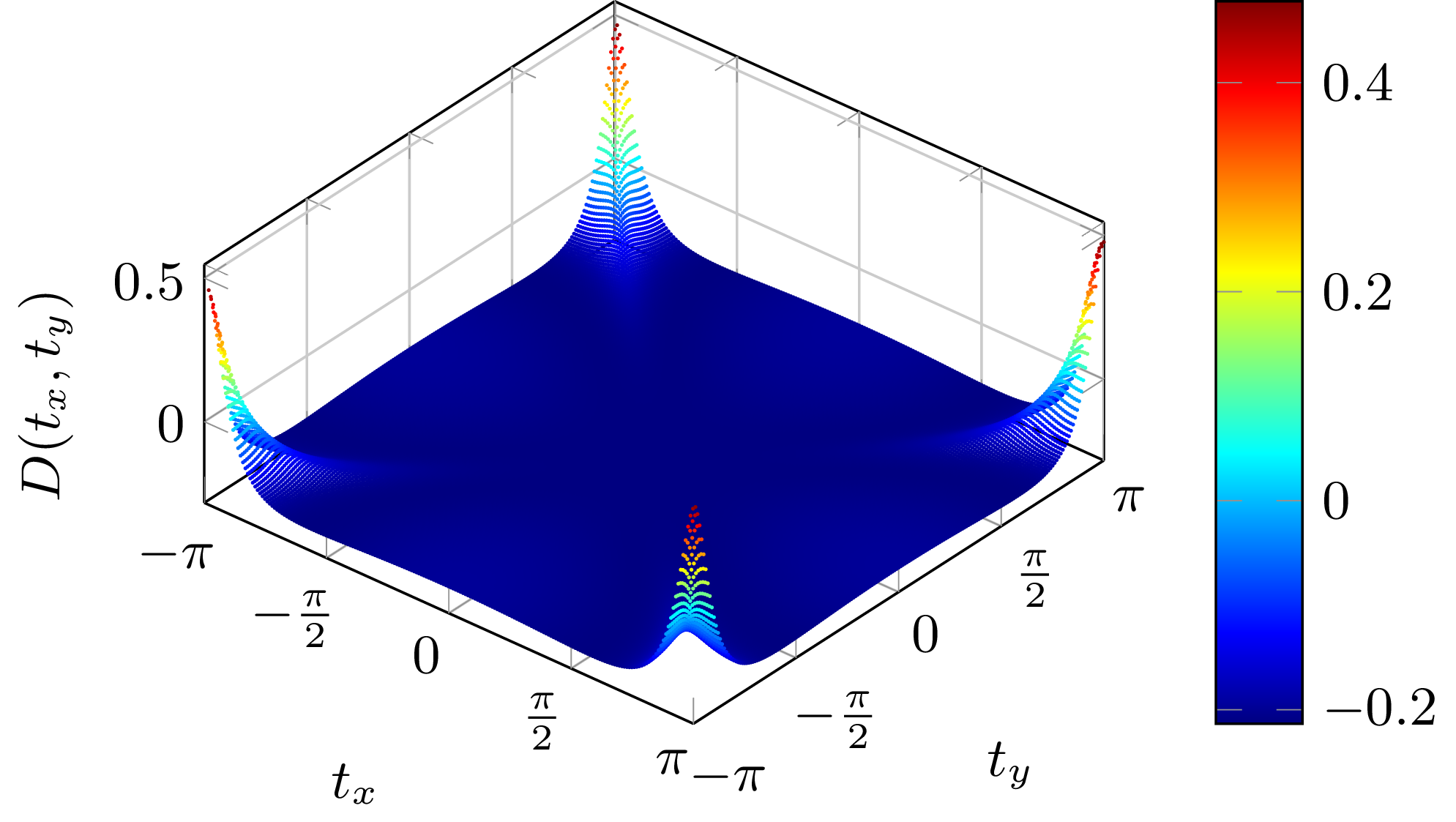}
\caption{Direct particle-hole vertex $D(\vec t)$.\label{fig:cph}}
\end{subfigure}
\begin{subfigure}{0.5 \textwidth}
\includegraphics[scale=1.0]{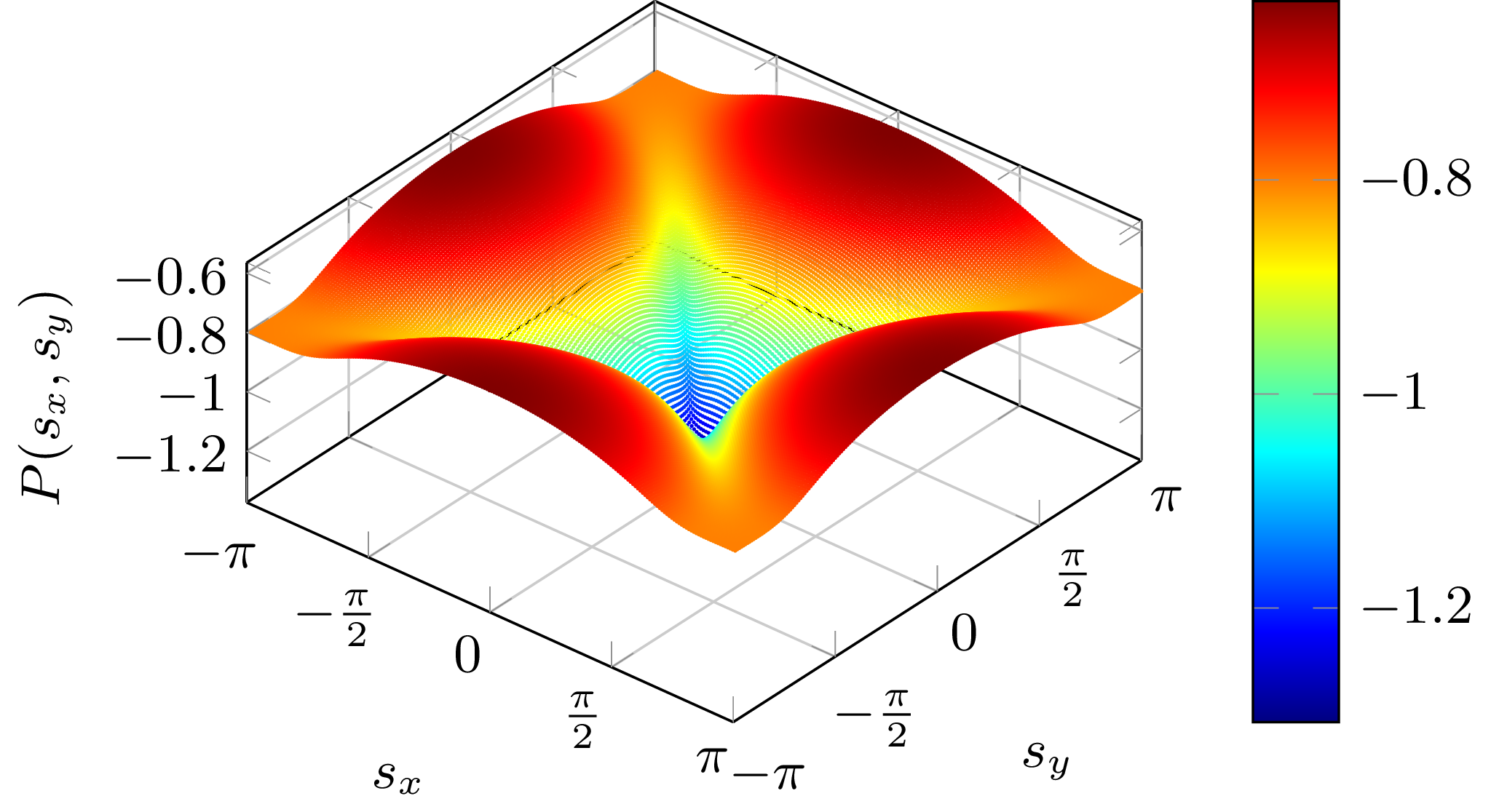}
\caption{Particle-particle vertex $P(\vec s)$. \label{fig:pp}}
\end{subfigure}
\caption{Projected vertices in the half-filled Hubbard model for parameters $U=2.0 \h t$, $T=0.1 \h t$, $N=200 \times 200$.
The vertices are plotted as functions of their respective main momentum, which ranges in the first Brillouin zone.}
\label{fig:phi}
\end{figure}


\begin{figure}[t]
\centering
\includegraphics[scale=1.0]{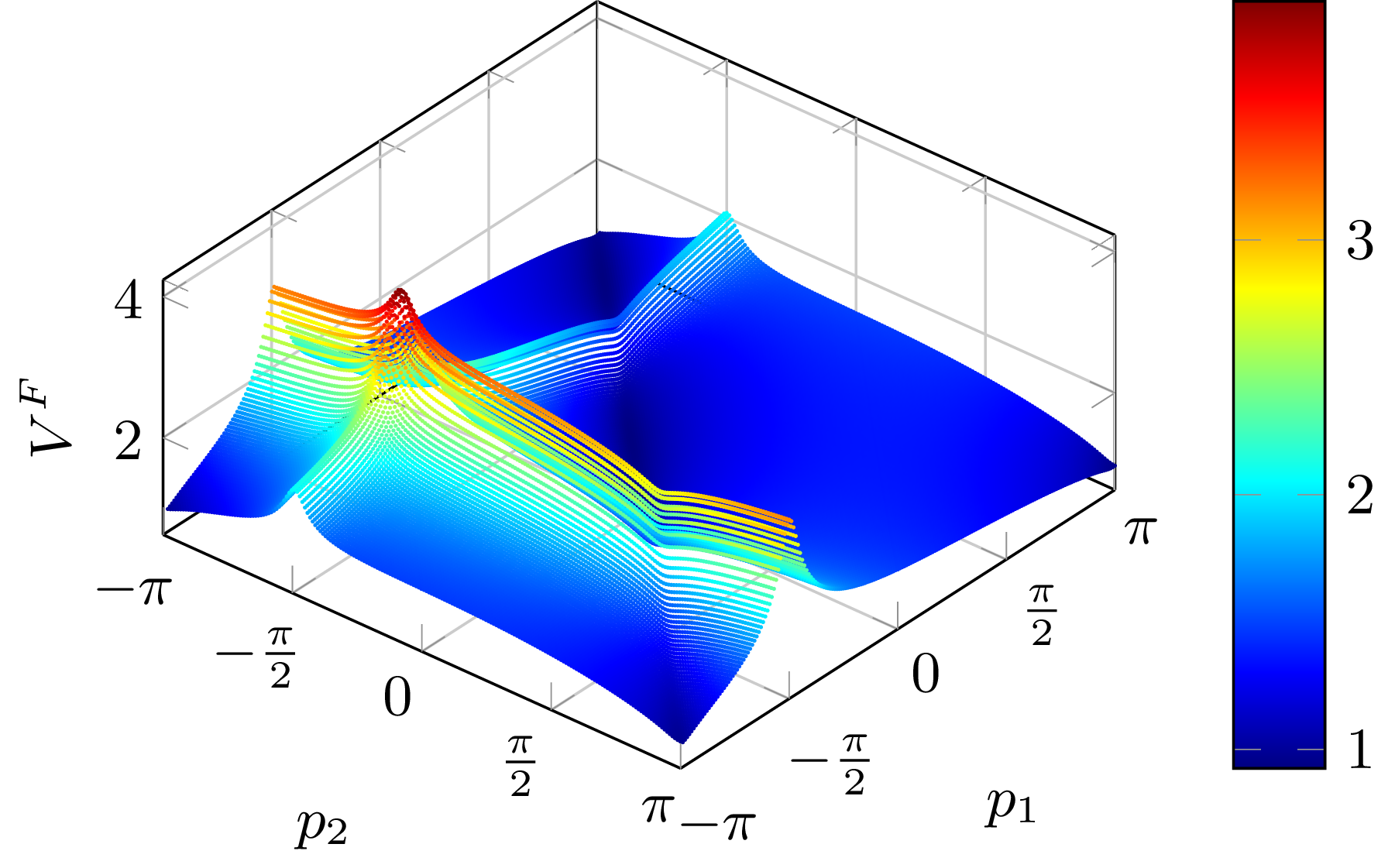}
\caption{Full vertex $V^{F}(\vec p_1, \vec p_2, \vec p_3)$ evaluated at $p_{1, x}=p_{1, y} =: p_1$, $p_{2, x}=p_{2, y} =: p_2$, and $p_{3, x} = p_{3, y} = \pi/2$ (same parameters as in Fig.~\ref{fig:phi}). \label{fig:F}}
\end{figure}

Finally, we present our numerical results for the Hubbard model at half filling and without next-nearest-neighbor hopping.
In this case, the non-interacting dispersion has the well-known fully nested Fermi surface, and the ground state at temperature $T=0$ and positive onsite interaction~$U$ should exhibit antiferromagnetic (AF) long-range order.
At nonzero temperatures, the Mermin-Wagner theorem prohibits long-range order, but as in the better-understood Heisenberg model there should still be longer-ranged AF correlations \cite{fazekas}.
A proper description of the state at $T>0$ with quantum many-body methods requires special care \cite{tremblay,katanintoschi}, and many methods like DCA or fRG replace the low-$T$ short-range ordered state with a long-range ordered state up to an artificial N\'eel temperature $T_{\mathrm c}$.
The major goal of the present analysis is to show that our channel-decomposed parquet scheme can reproduce the approach to long-range order with leading AF correlations, with a smaller remnant $T_{\mathrm c}$. 

The main observable that we study here is the full one-particle-irreducible vertex $F$.
Its momentum structure near the AF instability is well known from fRG studies and discussed e.g.~in Sct.~IIIB of Ref.~\onlinecite{metzner2012}.
Very close to the instability, which occurs at some nonzero $T_{\mathrm c}$ in the usual fRG approximations, the vertex has the leading momentum dependence 
\begin{equation}
V_{\mathrm{crit}} (\vec{p}_1,\vec{p}_2,\vec{p}_3) = \frac{J}{4} \h ( 2 \, \delta_{\vec{p}_1- \vec{p}_3, \h \vec{Q}} + \delta_{\vec{p}_3- \vec{p}_2, \h \vec{Q}}) \, , \label{vaf}
\end{equation}
with $J \propto 1/|T-T_{\mathrm c}|$. Here, $\vec{p}_1$ and $\vec{p}_2$ are incoming momenta, and $\vec{Q} = (\pi,\pi)$. 
This effective interaction can be transformed onto the real lattice, leading to an infinitely-long-ranged 
AF spin-spin interaction
\begin{equation}
J \h \sum_{\langle i, \h j \rangle } e^{i \vec{Q} \hh \cdot \hh (\vec{R}_i-\vec{R}_j)} \h \vec{S}_i \cdot \vec{S}_j \,, 
\end{equation}
with spin operators defined as
\begin{equation}
\vec{S}_{i} = \frac{1}{2} \h \sum_{\alpha, \h \beta} \vec{\sigma}_{\alpha \beta} \, c^\dagger_{i,\alpha} c_{i,\beta} \,.
\end{equation}
The same expression is also found in the random phase approximation (RPA) by using bare Green functions when ladder and bubble diagram chains are summed up (see Ref.~\onlinecite{scalapino93} Eq.~(5) close to the divergence).
The fRG changes the RPA results by reducing $T_{\mathrm c}$ and adding more non-divergent structure to the vertex.

\begin{figure*}[t]
\centering
\begin{subfigure}[t]{0.33 \textwidth}
\centering
\includegraphics[scale=1.0]{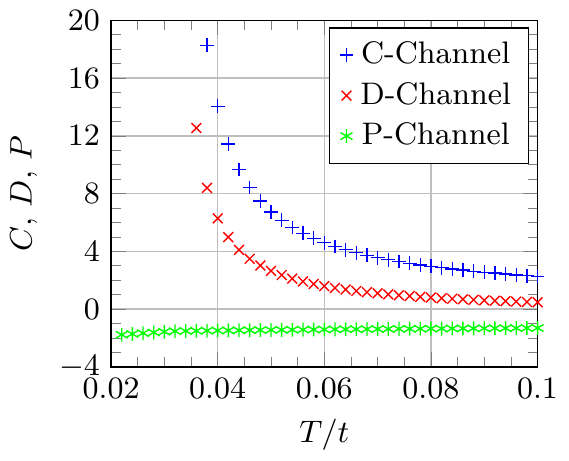}
\caption{$C(\pi, \pi)$, $D(\pi, \pi)$, and $P(0,0)$. \label{fig:temp}}
\captionsetup{justification=centering,margin=2cm}
\end{subfigure}%
\begin{subfigure}[t]{0.33 \textwidth}
\centering
\includegraphics[scale=1.0]{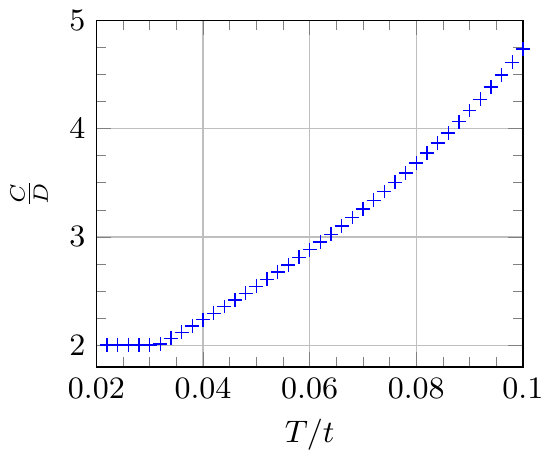}
%
\caption{$C(\pi,\pi)/D(\pi,\pi)$. \label{fig:CD}}
\captionsetup{justification=centering,margin=2cm}
\end{subfigure}%
\begin{subfigure}[t]{0.33 \textwidth}
\centering
\includegraphics[scale=1.0]{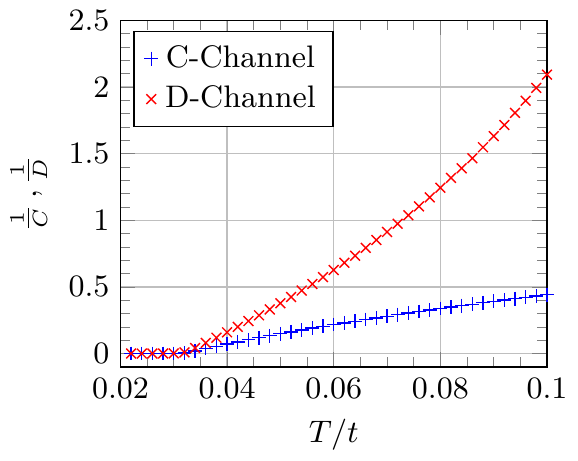}
\caption{$1/C(\pi, \pi)$ and $1/D(\pi,\pi)$. \label{fig:inv}}
\captionsetup{justification=centering,margin=2cm}
\end{subfigure}%
\caption{Extremum values of the projected vertices as functions fo the temperature $T$ for parameters $U = 2.0 \h t$ and $N = 200 \times 200$.}
\end{figure*}

Figures \ref{fig:ph}--\ref{fig:pp} show the numerical results for the vertices $D(\vec t)$, $C(\vec u)$, $P(\vec s)$ defined by Eqs.~\eqref{final_D}--\eqref{final_P} as functions of their respective main momentum, which ranges in the first Brillouin zone. For these computations an initial interaction of $U=2.0 \h t$ was used.
The temperature was set to a small value of $T = 0.1 \hh t$, and the computations were performed on a $200 \times 200$ grid.  We further mention that in order to improve the convergence, we have transformed the recursive formulas according to \cite{held_new,held2,jarrell_sym}
\begin{equation}
x_{i+1} = f(x_i) \quad \rightarrow \quad x_{i+1} = \alpha \h f(x_i) + (1-\alpha) \h x_{i}
\end{equation}
with a constant $\alpha \in (0,1]$, which in our implementation was set to $0.5$\hh. The evaluation of Eqs.~\eqref{eq:PA_projected_1}--\eqref{eq:PA_projected_3} takes only a few minutes on a laptop.

One can see that the crossed particle-hole vertex $C(\vec u)$ is always positive and strongly peaked at $\vec{u}= (\pi,\pi)$.
Similarly, the direct particle-hole vertex $D(\vec t)$ is peaked at $\vec{t}=(\pi,\pi)$,
but it has also negative contributions and its absolute value stays below that of $C(\vec u)$. The two strong peak features can be clearly associated with the two terms of Eq.~\eqref{vaf}, where the $\vec{t} = (\pi,\pi)$ feature is represented by the first term and the 
$\vec{u}= (\pi,\pi)$ peak by the second term.
By contrast, $P(\vec s)$ peaks at $\vec{s}=(0,0)$ and takes only negative values. This is also clearly understood and expected from the sign and momentum dependence of the particle-particle loop diagram. 
For completeness, we also show the full vertex
\begin{equation}
	\begin{aligned}
V^F(\vec p_1, \vec p_2, \vec p_3) = &U + V^{\mathrm{ph,d}}(\vec p_3 - \vec p_2)+\\
			&V^{\mathrm{ph,c}}(\vec p_1 - \vec p_3) + V^{\mathrm{pp}}(\vec p_1 + \vec p_2)
	\end{aligned}
\end{equation}
in Fig.~\ref{fig:F}.
By setting $\vec p_3 = (\pi/2, \pi/2)$, one clearly sees two peaked lines at $\vec p_1 = (-\pi/2, -\pi/2)$ and at $\vec p_2 = (-\pi/2, -\pi/2)$, which correspond to $\vec{t}=(\pi, \pi)$ and $\vec{u}=(-\pi,-\pi)$, respectively.
Furthermore, one observes a dip at $\vec{p_1}=-\vec{p_2}$, which results from the $V^{\mathrm{pp}}$ contribution.

When further decreasing the temperature, we find a divergence of $C$ and $D$ at $(\pi,\pi)$ as shown in Fig.~\ref{fig:temp}. The behavior close to the divergence is well approximated by
\begin{equation}
C(T) \sim \frac{1}{|T-T_{\mathrm c}|} \,,
\end{equation}
where $T_{\mathrm c}$ denotes the N\'eel critical temperature. In our implementation we find $T_{\mathrm c} \approx 0.035 \h t$
(at this value, the vertex is more than three times larger than the bandwidth).
In addition, as shown in Fig.~\ref{fig:CD}, we observe the relation $C /D \to 2 $ at $(\pi,\pi)$ and close to the divergence at $T_{\mathrm c}$, which is precisely the ratio between the two terms in Eq.~\eqref{vaf} thus reproducing the results from Ref.~\onlinecite{metzner2012}.
This also matches the behavior of previous RPA studies close to the divergence as seen for instance in Ref.~\onlinecite{scalapino93}. The parquet vertex in the TU approximation shows the same behavior of a nesting-driven AF ordering instability expected from RPA. 

\begin{figure}[t]
\begin{subfigure}{0.25 \textwidth}
\includegraphics[scale=1.0]{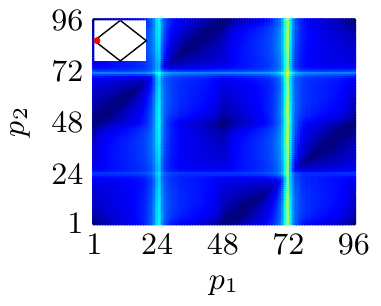}
\caption{ TU parquet -- $k_3$ at $1$. }
\end{subfigure}%
%
%
\begin{subfigure}{0.25 \textwidth}
\includegraphics[scale=1.0]{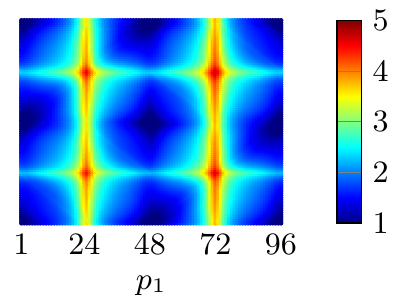}
\caption{ N-patch fRG -- $k_3$ at $1$. }
\end{subfigure}


\begin{subfigure}{0.25 \textwidth}
\includegraphics[scale=1.0]{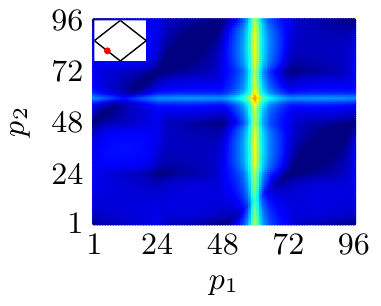}
\caption{ TU parquet -- $k_3$ at $13$. }
\end{subfigure}%
%
%
\begin{subfigure}{0.25 \textwidth}
\includegraphics[scale=1.0]{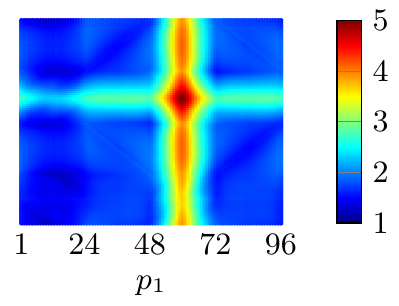}
\caption{ N-patch fRG -- $k_3$ at $13$. }
\end{subfigure}
\caption{Comparison of  TU parquet data (left plots) with ($N=96$)-patch fRG data (right plots) for the full vertex in the half-filled Hubbard model at $U=2t$. Colors encode the magnitude of the couplings.
The two incoming momentum indices $p_1$ and $p_2$ label 96 points on the Fermi surface indicated in the insets in the left plots.
Point 1 starts at $(-\pi,0)$, point 24 is near $(0,-\pi)$ and point 48 is near $(\pi,0)$.
The first outgoing momentum $p_3$ is taken to be at point 1 in the upper plots 
(see red bullet in the inset in the left upper plot)
or at point 13
(red bullet in the inset in the lower left plot).
For the parquet data $T=0.1\h t$ was used, while for the fRG a higher  $T=0.4\h t$ was chosen in order to achieve similar maximal values of the couplings.
The fRG data was obtained by the same code as in Refs.~\onlinecite{metzner2012} or \onlinecite{hsfr2001}.  
\label{fig:fRG_comp}}
\end{figure}

The TU parquet results for the Hubbard model also agree with fRG treatments of the same model as described e.g.~in Ref.~\onlinecite{metzner2012}.
In fact, the full parquet vertex constructed from the three channels can be directly compared with fRG data for the same model parameters, as shown e.g.\ in Fig.~10 of Ref.~\onlinecite{metzner2012}. There, $F$ has been obtained from an $N$-patch fRG scheme in the `standard' level-two truncation \cite{metzner2012}, also without self-energy corrections as in our work here. The comparison of our results with the fRG data obtained with the same code as in Ref.~\onlinecite{metzner2012} can be seen in Fig.~\ref{fig:fRG_comp}. The results for the full vertex match qualitatively, in terms of the enhancement features and also where the vertex remains small. The qualitative agreement with these fRG works supports the basic validity of our method. It also confirms the sensibility of the zeroth-order truncation in the form factor expansion for this case, as the analysis in Ref.~\onlinecite{metzner2012} does not rely on form factors. The parquet data also agree with a recent preprint, Ref.~\onlinecite{tagliavini}, that for the same parameters uses a form-factor expansion and at least partially includes non-local form factors. Hence, our method passes this qualitative sanity check.

The observable differences in Fig.~\ref{fig:fRG_comp} concern the resolution of the enhancement features. The parquet features are much narrower than the corresponding ones in fRG. One reason for this is that the fRG divergence scale $T^{\mathrm{fRG}}_{\mathrm c}\sim 0.11\h t$ for the approximation used is about three times higher than in parquet, and we had to choose $T=0.4\h t$ in order to find a similar magnitude of the largest couplings than in the parquet data shown here for $T=0.1 \h t$. Furthermore, the momentum resolution of the parquet result is given by the $200 \times 200$ grid points in the whole Brillouin zone, while the fRG works with just 96 points on the Fermi surface.  

The difference in the divergence scales should be discussed in more detail. The first point to keep in mind is that the true $T_{\mathrm c}$ for the AF ordering instability should be zero because of the Mermin-Wagner theorem. The self-consistent parquet approximation with the bare interaction as fully irreducible vertex, but including self-energy feedback on the internal lines, was argued\cite{bickers1992} to correctly describe the finite-$T$ correlations of Heisenberg-like order parameters, i.e., to fulfill the Mermin-Wagner theorem. In comparison with this, we remark that we do not include self-energies here because these would require a refined treatment of the frequency dependencies within our scheme (cf.~Refs.~\onlinecite{held_new, held2}). The same approximation holds for the $N$-patch data shown in Fig. \ref{fig:fRG_comp}. The parquet treatment includes more perturbative corrections than the fRG in the level-2 truncation, and indeed, the $T_{\mathrm c}$ found here is smaller, $\sim 0.035 \h t$ instead of $T^{\mathrm{fRG}}_{\mathrm c}\sim 0.11t$, as visible in Fig.~\ref{fig:inv}.
This nourishes hope that a parquet approach with self-energy feedback could actually get close to fulfillment of the Mermin-Wagner constraints.
Furthermore, in the recent preprint \onlinecite{tagliavini}, the multiloop-fRG scheme including frequency dependence of the interaction and self-energy feedback was applied to the same situation in the Hubbard model. The multiloop corrections reconstruct parquet contributions that are missed in the level-2 truncation of the fRG and result also in a reduced divergence scale compared to the previous fRG results.
The upshot of this comparison is that the quantitative picture at low $T$ depends on the further approximations used. The divergence scale becomes indeed smaller if a better approximation is used and should - theoretically, which may be hard in practice - reach zero if all parquet approximation terms and self-energies are included, and if sufficient momentum and frequency resolution is obtained. 
 Here, our numerically efficient parquet scheme may be a good starting point for further refinements.

\section{Conclusion and outlook}

We have presented a channel decomposition of the parquet equations, which effectively reduces the number of momentum arguments of the two-particle-reducible vertex functions. This method relies on introducing resolutions of unity in a form-factor basis, of which only a finite set of basis functions is kept. In this sense, the scheme can be called truncated-unity (TU) parquet approximation in analogy to the recently developed truncated-unity functional renormalization group (TUfRG) method \cite{julian,David17,Schober17}. In the TUfRG, the convergence in the number of form factors kept has turned out to be rather quick in most parameter regimes \cite{julian,David17}.

In a numerical implementation of the parquet equations, the channel decomposition effectively reduces the memory consumption from $\mathcal O(N^3)$ to $\mathcal O(N)$, where $N$ denotes the number of Bloch momenta taken into account. Furthermore, this method is particularly suitable for the parallelization on a large number of compute nodes. Since memory consumption is generally regarded as the ``bottle neck'' for implementing the parquet equations, the channel decomposition may allow one to reach a much higher precision in predicting ground-state orderings and critical scales in many-body models of condensed matter physics. 

To benchmark our method, we have implemented the TU parquet equations disregarding self-energy and frequency dependencies, and restricting attention to the lowest form factors. By means of the channel decomposition, we could study momentum resolutions with $\mathcal O(10^4)$ momenta in the first Brillouin zone with only a few minutes computing time on a standard laptop.
 Our results for the two-particle-reducible vertices in the fully nested Hubbard model on the square lattice qualitatively match those of previous fRG studies of the same model and get closer to a fulfillment of the Mermin-Wagner theorem than published fermionic fRG works. A more sophisticated implementation of the TU parquet equations which takes into account more form factors and/or the frequency dependencies of the vertices may improve this issue even more and is currently underway.

\begin{acknowledgments}
We thank K.\ Held, C.\ Hille, A.\ Kauch, F.\ Kugler, J.\ Lichtenstein, T.\ Reckling, M.\ Salmhofer, D.\ S\'{a}nchez de la Pe\~na, A.\ Tagliavini, D.\ Vilardi, and J.\ von Delft for helpful discussions.
We further acknowledge the computing time granted on the RWTH Compute Cluster. This work was supported by the DFG grants HO 2422/10-1, 11-1, and 12-1, DFG RTG 1995, and by the Vienna Computational Materials Laboratory (FWF SFB ViCoM F41).
\end{acknowledgments}

\appendix
\begin{widetext}
\section{SU(2)-symmetric parquet equations} \label{appendix:spin}

In this appendix, we show how spin-SU(2)-symmetry can be used to facilitate the evaluation of the parquet equations.
In the SU(2)-symmetric case, all vertices effectively depend on only three spin arguments, the fourth being determined by spin conservation.
Correspondingly, in the parquet equations given in Ref.~\onlinecite{held},
the sum over internal spin indices can be limited to only a few spin configurations. 
Often, one also considers superpositions of such configurations to evaluate the parquet equations \cite{bickers, held}. In this article, we use instead the following decomposition of vertices, which generally holds in the SU(2)-symmetric case (see e.g.~Ref.~\onlinecite{SH00}):
\begin{equation}
\varGamma_{\sigma_1 \sigma_2\sigma_3\sigma_4}(k_1,k_2,k_3,k_4) = -V^{\varGamma}(k_1,k_2,k_3,k_4) \, \delta_{\sigma_1 \sigma_4} \hh \delta_{\sigma_2 \sigma_3} + W^{\varGamma}(k_1,k_2,k_4,k_3) \,\delta_{\sigma_1 \sigma_3} \hh \delta_{\sigma_2 \sigma_4} \,.
\label{eq:gen-su2}
\end{equation}
For a general $\varGamma$, the component functions $V^\varGamma$ and $W^\varGamma$ can be independent of each other. However, if we consider the full vertex, $\varGamma \equiv F$, we can further employ its antisymmetry under the simultaneous interchange of two momentum and spin arguments, i.e., the so-called crossing relation \cite{bickers}
\begin{equation}
F_{\sigma_1 \sigma_2\sigma_3\sigma_4}(k_1,k_2,k_3,k_4) = -F_{\sigma_1 \sigma_2\sigma_4\sigma_3}(k_1,k_2,k_4,k_3) \,.
\label{eq:relF}
\end{equation}
From this equation, one can derive the following relation between the coefficient functions in Eq.~\eqref{eq:gen-su2}:
\begin{equation}
V^F(k_1, k_2, k_3, k_4) = -F_{\uparrow \downarrow \downarrow \uparrow}(k_1, k_2, k_3, k_4) = F_{\uparrow \downarrow \uparrow \downarrow}(k_1, k_2, k_4, k_3) = W^F(k_1, k_2, k_3, k_4) \,, \label{eq:Fsu2}
\end{equation}
hence $V^F \equiv W^F$. Similarly, the particle-particle vertex $\varPhi_{\mathrm{pp}}$ is also antisymmetric under the exchange of two momentum and spin arguments, hence Eqs.~\eqref{eq:relF} and \eqref{eq:Fsu2} hold analogously for the particle-particle vertex. Another similar case is the bare interaction of the Hubbard model, which is given by
\begin{equation}\label{ini_int}
F^0_{\sigma_1 \sigma_2 \sigma_3 \sigma_4}(k_1, k_2, k_3, k_4) = U \h\big( \delta_{\sigma_1 \sigma_3} \hh \delta_{\sigma_2 \sigma_4} - \delta_{\sigma_1 \sigma_4} \hh \delta_{\sigma_2 \sigma_3} \big) \,.
\end{equation}
On the other hand, for the particle-hole vertices the relations between the $V$- and $W$-functions are different. For these functions, we can use another crossing relation \cite{bickers}, namely
\begin{equation}
\varPhi^{\mathrm{ph,d}}_{\sigma_1\sigma_2\sigma_3\sigma_4}(k_1,k_2,k_3,k_4) = -\varPhi^{\mathrm{ph,c}}_{\sigma_1 \sigma_2 \sigma_4 \sigma_3}(k_1,k_2,k_4,k_3) \,.
\label{eq:cross}
\end{equation}
With this, we obtain the following identities:
\begin{align} \label{eq:Phisu1}
V^{\mathrm{ph,d}} \equiv W^{\mathrm{ph,c}} \,, \qquad \textnormal{and} \quad
V^{\mathrm{ph,c}} \equiv W^{\mathrm{ph,d}} \,,
\end{align}
where we have abbreviated \mbox{$V^{\mathrm{ph,d}} \equiv V^{\varPhi^{\mathrm{ph,d}}}$,} etc. Together, these relations imply Eqs.~\eqref{eq_cross_1}--\eqref{eq_cross_3} in the main text.

Next, one can reformulate the parquet equations \eqref{eq:parquet1_full}--\eqref{eq:parquet3_full} in terms of the spin-independent $V$-functions. This calculation is analogous to the derivation of the SU(2)-symmetric RG equations in Ref.~\onlinecite{SH00}, and hence we only state the result here (cf.~also Ref.~\onlinecite[Eqs.~(2)--(4)]{julian}):
\begin{align}
 V^{\mathrm{ph,d}}(p_1, p_2, p_3) & = \frac{k_{\mathrm B} T}{N} \sum_k G(k) \, G(k + p_2 - p_3) \\[1pt] \nonumber
 & \quad \, \times \Big( \h 2 \h V^F(p_1, k + p_2 - p_3, k) \, [V^F - V^{\mathrm{ph,d}}](k, p_2, p_3) \\[2pt] \nonumber
 & \hspace{1cm} - V^F(p_1, k + p_2 - p_3, k) \, [V^F - V^{\mathrm{ph,c}}](k, p_2, k + p_2 - p_3) \\[2pt] \nonumber
 & \hspace{1cm} - V^F(p_1, k + p_2 - p_3, \h p_1 + p_2 - p_3) \, [V^F - V^{\mathrm{ph,d}}](k, p_2, p_3) \h \Big) \,, \\[8pt]
 V^{\mathrm{ph,c}}(p_1, p_2, p_3) & = -\frac{k_{\mathrm B} T}{N} \sum_k G(k) \, G(k + p_3 - p_1) \, V^F(p_1, k + p_3 - p_1, p_3) \, [ V^F - V^{\mathrm{ph,c}}](k, p_2, k + p_3 - p_1) \,, \\[5pt]
 V^{\mathrm{pp}}(p_1, p_2, p_3) & = -\frac{k_{\mathrm B} T}{N} \sum_k G(k) \, G(p_1 + p_2 - k) \, V^F(p_1, p_2, k) \, [ V^F - V^{\mathrm{pp}} ](p_1 + p_2 - k, k, p_3) \,.
\end{align}
The full vertex $F$ is correspondingly given by
\begin{equation}
V^{F} = U + V^{\mathrm{ph,d}} + V^{\mathrm{ph,c}} + V^{\mathrm{pp}} \,.
\label{eq:VF}
\end{equation}
Furthermore, these SU(2)-symmetric parquet equations can be projected onto the various channels defined in the main text. Performing the same steps as in the derivation of Eqs.~\eqref{eq:projected1}--\eqref{eq:projected3} for the spin-dependent vertices, we arrive at the following TU parquet equations for the spin-independent vertices (cf.~Ref.~\onlinecite[Eqs.~(22)--(24)]{julian}):
\begin{align}
 \hat D[V^{\mathrm{ph,d}}]_{\ell_1 \ell_2}(t) & = \sum_{\ell, \h \ell'} \Big( 2 \, \hat D[V^F]_{\ell_1 \ell}\h(t) \,\hh L^{\mathrm{ph}}_{\ell \ell'}(t) \,\hh \hat D[V^F - V^{\mathrm{ph,d}}]_{\ell' \ell_2}\mh(t) \\[-5pt] \nonumber
 & \hspace{1.2cm} - \hat D[V^F]_{\ell_1 \ell}\h(t) \,\hh L^{\mathrm{ph}}_{\ell \ell'}(t) \,\hh \hat C[V^F - V^{\mathrm{ph,c}}]_{\ell' \ell_2}\mh(t) \\[5pt] \nonumber
 & \hspace{1.2cm} - \hat C[V^F]_{\ell_1 \ell}\h(t) \,\hh L^{\mathrm{ph}}_{\ell \ell'}(t) \,\hh \hat D[V^F - V^{\mathrm{ph,d}}]_{\ell' \ell_2}\mh(t) \Big) \,, \\[10pt]
 \hat C[V^{\mathrm{ph,c}}]_{\ell_1 \ell_2}(u) & = -\sum_{\ell, \h \ell'} \hat C[V^F]_{\ell_1 \ell}\h(u) \,\hh L^{\mathrm{ph}}_{\ell \ell'}(u) \,\hh \hat C[V^F - V^{\mathrm{ph,c}}]_{\ell' \ell_2}\mh(u) \,, \\[5pt]
 \hat P[V^{\mathrm{pp}}]_{\ell_1 \ell_2}(s) & = - \sum_{\ell, \h \ell'} \hat P[V^F]_{\ell_1 \ell}(s) \,\hh L^{\mathrm{pp}}_{\ell \ell'}(s) \,\hh \hat P[V^F - V^{\mathrm{pp}}]_{\ell' \ell_2}(s) \,,
\end{align}
where the loop terms are again given by Eqs.~\eqref{loop_ph}--\eqref{loop_pp}. Thus, we have shown that the crossing relations, i.e., the antisymmetry of $\varPhi_{\mathrm{pp}}$ and the initial interaction, as well as Eq.\ \eqref{eq:cross}, allow one to express all spin-dependent vertices in terms of the spin-independent $V$-functions and thereby to reduce the number of functions one has to keep track off.
We note that other implementations of the parquet equations such as Refs.~\onlinecite{held_new,jarrell_sym} do not exploit these crossing relations explicitly but
instead enforce them during the iteration process to improve the convergence.

\end{widetext}

\bibliography{thebib}

\end{document}